\documentclass{revtex4-2}

\usepackage[a4paper, total={6.7in, 8.5in}]{geometry}
\usepackage[utf8]{inputenc}
\usepackage{graphicx}
\usepackage{subcaption}

\usepackage{newpxmath,newpxtext}
\usepackage{graphicx}
\usepackage{epsfig}
\usepackage{tensor}
\usepackage{hyperref}
  
\newcommand{\pder}[2]{\dfrac{\partial#1}{\partial#2}}
\newcommand{\pdder}[3]{\dfrac{\partial^2 #1}{\partial #2 \partial #3}}
\newcommand{\dder}[2]{\dfrac{\delta#1}{\delta#2}}
\newcommand{\Gd}{\mathcal{G}}
\newcommand{\R}{\mathcal{R}}
\newcommand{\de}{\mathrm{d}}

\begin{document}

\title{Reissner-Nordström and Kerr-like solutions in Finsler-Randers Gravity}

\author{Georgios Miliaresis}
\email{s06gmili@uni-bonn.de}
\affiliation{Department of Physics and Astronomy, Rheinische Friedrich-Wilhelms-Universität Bonn,
Nußallee 12, 53115 Bonn, Germany}

\author{Konstantinos Topaloglou}
\email{konstanti.topaloglou@studio.unibo.it}
\affiliation{Via Angelo Venturoli 39, 40138 Bologna, IT}

\author{Ioannis Ampazis}
\email{johnampazis@gmail.com}
\affiliation{Konstantinoupoleos 77, Pefki, Athens, Greece}

\author{Nefeli Androulaki}
\email{mynameisnefeli@gmail.com}
\affiliation{University of Patras, Department of Mathematics, 26504 Rion, Greece}

\author{Emmanuel Kapsabelis}
\email{manoliskapsabelis@yahoo.gr}
\affiliation{Section of Astrophysics, Astronomy and Mechanics, Department of Physics, National and
Kapodistrian University of Athens, Panepistimiopolis, 15784 Athens, Greece}

\author{Emmanuel N. Saridakis}
\email{msaridak@noa.gr}
\affiliation{National Observatory of Athens, Lofos Nymfon, 11852 Athens, Greece}
\affiliation{CAS Key Laboratory for Research in Galaxies and Cosmology, School of Astronomy and Space Science, University of Science and Technology of China, Hefei 230026, China}
\affiliation{Departamento de Matem\'{a}ticas, Universidad Cat\'{o}lica del Norte, Avda. Angamos 0610, Casilla 1280, Antofagasta, Chile}

\author{Panayiotis C. Stavrinos}
\email{pstavrin@math.uoa.gr}
\affiliation{Department of Mathematics, National and Kapodistrian University of Athens,
Panepistimiopolis, 15784 Athens, Greece}

\author{Alkiviadis Triantafyllopoulos}
\email{alktrian@phys.uoa.gr}
\affiliation{Section of Astrophysics, Astronomy and Mechanics, Department of Physics, National and
Kapodistrian University of Athens, Panepistimiopolis, 15784 Athens, Greece}

\begin{abstract}
In a previous study  we investigated the spherically symmetric Schwarzschild and
Schwarzschild–de Sitter solutions within a Finsler–Randers-type geometry. In this work we extend our analysis to charged and rotating
solutions, focusing on the Reissner–Nordström and Kerr-like metrics in the Finsler–Randers gravitational framework. In particular, we extract the modified gravitational field equations and we examine the geodesic equations, analyzing particle trajectories and quantifying the deviations from their standard counterparts. Moreover, we compare the results  with the predictions of general relativity, and we discuss how   potential deviations from  Riemannian geometry could be reached observationally.

\end{abstract}

\maketitle


\section{Introduction}
\begin{figure}[h]
  \centering
  \includegraphics[width=1\textwidth]{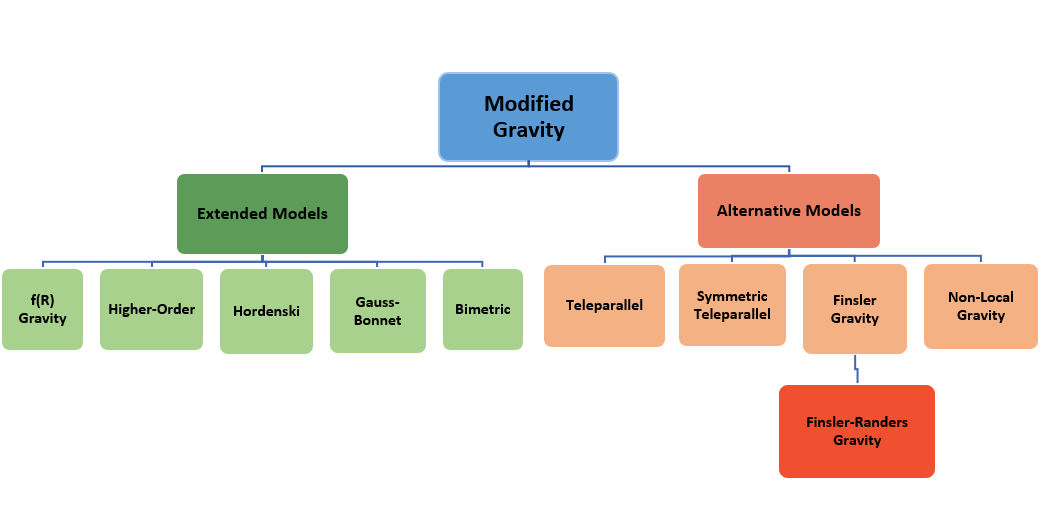} 
  \caption{Schematic summary of some classes of modified gravity theories, including Finsler-Randers as an innovative extension of Finsler gravity.}
  \label{fig:my_image}
\end{figure}

General Relativity (GR) has been remarkably successful in describing gravity, black holes, and cosmic expansion, based on Riemannian geometry, which is a homogenous and isotropic model. However, the nature of dark energy, responsible for the accelerated expansion of the universe, and dark matter, inferred from galactic dynamics, remain unresolved. Additionally, GR  faces singularities issues, while it  is non-renormalizable,
which motivates the search for alternative theories. Modified gravity theories are obtained through extensions and modifications  of general relativity  \cite{CANTATA:2021asi}. These theories are crucial to modern cosmology, offering key insights and a foundational framework for interpreting the Universe's physical phenomena \cite{DiValentino:2025sru} (See figure \ref{fig:my_image} for a schematic summary of some classes of modified gravity theories).

\vspace{2cm}
On the other hand, observations of the Cosmic Microwave Background (CMB) reveal that the universe is not perfectly isotropic. On local scales, structures such as galaxies, clusters, voids, and cosmic filaments introduce local anisotropies, challenging the assumption of large-scale homogeneity and symmetry. In addition, a directional asymmetry of mass particles has recently been investigated \cite{Shao2024}. Moreover, cosmic expansion may exhibit directional dependence, influenced by factors such as primordial magnetic fields or asymmetric galaxy rotation \cite{shamir2025}. Extra-dimensional theories extend GR by introducing additional degrees of freedom, offering alternative explanations for cosmic evolution, gravity, and high-energy phenomena. The anisotropy observed in the universe may be directly related to its matter content \cite{MTW}. Specifically, it is considered that anisotropies emerging during the cosmic expansion could have contributed to the generation of particle creation, through mechanisms activated under conditions of rapid spacetime variation causing an anisotropic dark energy. In addition, anisotropy in cosmological models may provide insight into dark matter and the large-scale structure of the universe. Advances in observational and computational techniques may offer new ways to test these theories and refine our understanding of fundamental forces and spacetime structure.

 Anisotropy can also be produced by the interaction between the internal and external structures of spacetime. There are differences between internal and external anisotropy, with a vector field playing a key role in influencing the metric, for instance, structure of Riemann spacetime. A primordial vector field, such as a magnetic field, can be embedded within the structure of spacetime and cause anisotropic effects,influencing the event horizon. Another investigation with a different perspective at the papers \cite{campanelli2011,campanelli2006,campanelli2007}
proposes an ellipsoidal universe model to explain cosmic anisotropies, linking CMB
quadrupole anomalies and cosmic parallax to intrinsic geometric properties rather than
observational artifacts. In \cite{shamir2025} a significant rotational asymmetry is uncovered in
galaxies observed by the Advanced Deep Extragalactic Survey, suggesting early-universe
structure could influence galaxy formation and contribute to the Hubble tension. The authors in \cite{migkas2021} detect variation in $H_0$, pointing to large-scale anisotropies that challenge the assumption of cosmic isotropy. In \cite{pandya2021} this finding using galaxy cluster velocity dispersion
scaling relations, ruling out temperature measurement biases as an explanation. In \cite{tedesco2024}, a Bianchi anisotropic universe is explored, deriving a cosmic shear expression in terms of
eccentricity, offering a powerful approach to understanding deviations from homogeneity.
In this paper, cosmic anisotropy is investigated using a Bianchi model, highlighting the
connections between cosmic shear, eccentricity, and CMB quadrupole anomalies. It
emphasizes the importance of considering both anisotropy and inhomogeneity for a
comprehensive cosmological framework.

A promising geometric framework is Finsler geometry, which
extends Riemannian geometry by incorporating spacetime anisotropy and the gravitational
interactions depend on both position and direction. In recent years, significant
research has explored the physical and cosmological implications of Finsler geometry in different ways, investigating its potential applications to fundamental problems such as the Hubble tension, dark matter, dark energy \cite{triantafyllopoulos2020,kapsabelis2021,Stavrinos2004,Vacaru2005,Kouretsis2009,Skakala2011,Kostelecky2011,Vacaru2012,Pfeifer2012,Basilakos:2013hua,Elizalde2015,Pfeifer2011,Caponio2016,Brody2016,Hohmann2017,WangEPJC2017,Ellis2017,Silva2016,Wang2017,Shah2018,Fuster2018,Triantafyllopoulos2018,
Chaubey2019,Minas:2019urp,Bubuianu2019,Ikeda:2019ckp,HohmannU2020,Caponio2020,Relancio2020,HohmannEPJC2020,HohmannPRD2020,Javaloyes2021,Konitopoulos:2021eav,vanVoorthuizen2021,Hama2021,Stavrinos2021,Hama2022,Hama2023,Heefer2023,Narasimhamurthy2024,AnnamáriaFriedl-Szász2025} and with Lorentz-violation
\cite{Kostelecky2011,Kouretsis2009,Zhu2023,Savvopoulos2023,Villasenor2024}. Finslerian gravitational theories play a significant role by introducing corrections to the Einstein field equations and the Friedmann–Robertson–Walker cosmological equations, particularly in the context of anisotropic cosmological phenomena. These issues can be studied within the framework of Finsler or Finsler-like cosmology, offering alternative approaches to their resolutions. Furthermore, we present some works relevant to these applications, which have contributed to the developement of this investigation in a different perspective.

In Finslerian geometry, geodesics follow a modified variational principle, leading to
equations of motion with anisotropic corrections. These modifications influence
astrophysical phenomena such as deviations from standard GR predictions, the motion of
test particles, and the structure of black hole accretion disks. By deriving solutions within the
Finsler-Randers framework, researchers  obtain information about the role of anisotropic
effects in black hole physics, cosmic acceleration, and the broader implications for modified
gravity theories. Understanding these effects may play a crucial role in refining our
knowledge of the fundamental structure of spacetime and its evolution.

In this context, Finsler-Randers geometry is a special case of Finsler geometry incorporating an additional vector field,which provides a framework for modeling anisotropic gravitational interactions, where the gravitational field is described in the metric of the tangent bundle of spacetime. Finsler–Randers (FR) space has been introduced by G. Randers in 1941 \cite{Randers1941} as a Finslerian extension of Riemannian geometry, motivated by its applications in general relativity and electromagnetic theory
\cite{Brody2016,Ellis2017,Silva2016,Wang2017,Gibbons2007,Chaubey2018,Mavromatos2013,Basilakos2013,Vacaru2014,Vacaru2017,Caponio2018,Stavrinos2018,Triantafyllopoulos2018,Heefer2020,Elbistan2020,Raushan2020,TriantafyllopoulosEPJC2020,Silva2021,Angit2022,Lou2022,Voicu2023,Feng2023,Das2023,Kapsabelis2022,Kapsabelis2024,Chanda2024,El-Nabulsi2024}.
This model incorporates both forms of anisotropy, providing framework for understanding deviations from isotropy. 
The correlation between an anisotropic scalar parameter comes from the structure of Finsler-Randers spacetime with the dark energy, is key feature which modifies standard gravitational dynamics. This metric has also been introduced in cosmology by \cite{Stavrinos2008}.

In an alternative way, the anisotropy has been studied in the Schwarzschild – Finsler – Randers model (SFR) and it explores modifications to gravity through Finsler–Randers geometry, focusing on Schwarzschild-like solutions and their astrophysical implications \cite{pandya2021,tedesco2024,triantafyllopoulos2020}. These works examine anisotropies on a cosmological scale. In addition we mention some other studies \cite{Silagadze2010,Li2020,Kapsabelis2021,Ke-JianHe2024,Dehkordi2025,Yao2025}.

In this work, we extend Reissner-Nordström solution, which describes a charged, non-rotating black hole, as well as the Kerr solution, which describes the spacetime around a rotating, uncharged mass in general relativity, derived by R. Kerr in 1963 \cite{Kerr1963}. It extends the Schwarzschild solution by incorporating the source's angular momentum, and is therefore essential in modeling rotating black holes, featuring an event horizon and ergosphere which have important astrophysical implications. Modifications in these features of the black holes' spacetime provide deeper insights into black hole thermodynamics, stability, and gravitational lensing, highlighting the role of locally anisotropic geometries in gravitational and electromagnetic phenomena. Considerations in this generalized framework for R-N and Kerr with Finsler geometry have been done also in \cite{Rajpoot2015,Li2018,Rarras2024,Konoplya:2006br} and  \cite{Kerr1963,Kerr1970,Konoplya:2006br,Stavrinos2008,Visser2007,Cvetic2014,Vacaru2017,Kerr2023,Rarras2024,Dehkordi2025}.

The paper is organised as follows: in Section 2 the geometric structure of the Reissner–Nordström and Kerr spacetime models. In Section 3, we study the extension of the Finsler Reissner–Nordström spacetime, derive the field equations and their analytical solutions, and examine the behavior of geodesics and trajectories. In Section 4, we explore a solution for a Kerr-like spacetime, solving the field equations analytically for the polar and radial components, and  numerically for the time and azimuthal components of the anisotropy-inducing one-form. We also analyze the corresponding geodesics and trajectories, and additionally discuss the physical implications of the results and illustrate modified trajectories with figures. Furthermore, we briefly address the structure of the black hole horizon in Kerr-like spacetime. Finally, in Section 5 we summarize all findings in the conclusions.

\section{Geometrical structure of the models}
	In this section, we review the geometric and physical framework upon which we build our models. 
	The geometrical structure which describes our generalized space is a Lorentz tangent bundle $TM$, i.e. an 8-dimensional tangent bundle of a 4-dimensional manifold with local coordinates $\{x^\mu,y^\alpha\}$ where the indices of the $x$ variables are $\kappa,\lambda,\mu,\nu,\ldots = 0,\ldots,3$ and the indices of the $y$ variables are $\alpha,\beta,\ldots,\theta = 4,\ldots,7$.
	At each point on $TM$ there is defined an 8-dimensional tangent space, in which we make a special choice of basis which is known as adapted basis. It is defined as $\{E_A\} = \,\{\delta_\mu,\partial_\alpha\} $ with
	\begin{equation}
		\delta_\mu = \dfrac{\delta}{\delta x^\mu}= \pder{}{x^\mu} - N^\alpha_\mu(x,y)\pder{}{y^\alpha} \label{delta x}
	\end{equation}
	and
	\begin{equation}
		\partial_\alpha = \pder{}{y^\alpha}
	\end{equation}
	where $N^\alpha_\mu$ are the components of the nonlinear connection on $TM$.
	
	The nonlinear connection induces a split of the total space $TTM$ into a horizontal distribution $T_HTM$ and a vertical distribution $T_VTM$. This split is expressed as the Whitney sum:
	\begin{equation}
		TTM = T_HTM \oplus T_VTM .
	\end{equation}
	
	The geometry in hand is equipped with a Sasaki-type metric $\Gd$ on $TM$
	\begin{equation}
		\mathcal{G} = g_{\mu\nu}(x,y)\,\mathrm{d}x^\mu \otimes \mathrm{d}x^\nu + v_{\alpha\beta}(x,y)\,\delta y^\alpha \otimes \delta y^\beta  .\label{bundle metric}
	\end{equation}
	This metric is clearly split into a horizontal and a vertical one. For $TM$ to be Lorentzian, the individual metrics $g_{\mu\nu}$ and $v_{\alpha\beta}$ need to be pseudo-Finslerian.
	
	A pseudo-Finslerian metric $ f_{\alpha\beta}(x,y) $ is defined as one that has a Lorentzian signature of $(+,-,-,-)$ and that also obeys the following form:
	\begin{align}
		f_{\alpha\beta}(x,y) = \pm\frac{1}{2}\pdder{F^2}{y^\alpha}{y^\beta} \label{Fg}
	\end{align}
	where the {\it Finsler metric function} $F$ satisfies the following conditions:
	\begin{enumerate}
		\item $F$ is continuous on $TM$ and smooth on  $ \widetilde{TM}\equiv TM\setminus \{0\} $ i.e. the tangent bundle minus the null set $ \{(x,y)\in TM | F(x,y)=0\}$ . \label{finsler_field_of_definition}
		\item $ F $ is positively homogeneous of first degree on its second argument:
		\begin{equation}
			F(x^\mu,ky^\alpha) = kF(x^\mu,y^\alpha), \qquad k>0 .\label{finsler_homogeneity}
		\end{equation}
		\item The form 
		\begin{equation}
			f_{\alpha\beta}(x,y) = \pm\dfrac{1}{2}\pdder{F^2}{y^\alpha}{y^\beta} \label{finsler_metric} 
		\end{equation}
		defines a non-degenerate matrix: \label{finsler_nondegeneracy}
		\begin{equation}
			\det\left[f_{\alpha\beta}\right] \neq 0 .\label{finsler_nondegenerate}
		\end{equation}
	\end{enumerate}
	
	\subsection{Linear connection, curvature and torsion}
	In this work, we consider a $d-$connection $D$ on $TM$. This is a linear connection with coefficients $\{\Gamma^A_{BC}\} = \{L^\mu_{\nu\kappa}, L^\alpha_{\beta\kappa}, C^\mu_{\nu\gamma}, C^\alpha_{\beta\gamma} \} $ with non-vanishing components:
	\begin{align}
		{D_{\delta_\kappa}\delta_\nu = L^\mu_{\nu\kappa}(x,y)\delta_\mu} \quad &\quad D_{\partial_{\gamma}}\delta_\nu = C^\mu_{\nu\gamma}(x,y)\delta_\mu \label{D delta nu} \\
		{D_{\delta_\kappa}\partial_{\beta} = L^\alpha_{\beta\kappa}(x,y)\partial_{\alpha}} \quad & \quad D_{\partial_{\gamma}}\partial_{\beta} = C^\alpha_{\beta\gamma}(x,y)\partial_{\alpha} .\label{D partial b}
	\end{align}
	
	The $d-$connection is metric-compatible when the following conditions are met:
	\begin{equation}
		D_\kappa\, g_{\mu\nu} = 0, \quad D_\kappa\, v_{\alpha\beta} = 0, \quad D_\gamma\, g_{\mu\nu} = 0, \quad D_\gamma\, v_{\alpha\beta} = 0.
	\end{equation}
	A $d-$connection is uniquely defined when imposing the additional conditions:
	\begin{itemize}
		\item The $d-$connection is metric compatible
		\item Coefficients $L^\mu_{\nu\kappa}, L^\alpha_{\beta\kappa}, C^\mu_{\nu\gamma}, C^\alpha_{\beta\gamma} $ depend solely on the quantities $g_{\mu\nu}$, $v_{\alpha\beta}$ and $N^\alpha_\mu$
		\item Coefficients $L^\mu_{\kappa\nu}$ and $ C^\alpha_{\beta\gamma} $ are symmetric on the lower indices, i.e.  $L^\mu_{[\kappa\nu]} = C^\alpha_{[\beta\gamma]} = 0$
	\end{itemize}
	We use the symbol $\mathcal D$ instead of $D$ for a connection satisfying the above conditions, and call it a canonical and distinguished $d-$connection.
	The nonzero components of this connection are
	\begin{align}
		L^\mu_{\nu\kappa} & = \frac{1}{2}g^{\mu\rho}\left(\delta_kg_{\rho\nu} + \delta_\nu g_{\rho\kappa} - \delta_\rho g_{\nu\kappa}\right) \label{metric d-connection 1}  \\
		L^\alpha_{\beta\kappa} & = {\partial}_\beta N^\alpha_\kappa + \frac{1}{2}v^{\alpha\gamma}\left(\delta_\kappa v_{\beta\gamma} - v_{\delta\gamma}\,{\partial}_\beta N^\delta_\kappa - v_{\beta\delta}\,{\partial}_\gamma N^\delta_\kappa\right) \label{metric d-connection 2}  \\
		C^\mu_{\nu\gamma} & = \frac{1}{2}g^{\mu\rho}{\partial}_\gamma g_{\rho\nu} \label{metric d-connection 3} \\
		C^\alpha_{\beta\gamma} & = \frac{1}{2}v^{\alpha\delta}\left({\partial}_\gamma v_{\delta\beta} + {\partial}_\beta v_{\delta\gamma} - {\partial}_\delta v_{\beta\gamma}\right) .\label{metric d-connection 4}
	\end{align}
	The generalized Ricci scalar curvature is defined as:
	\begin{equation}
		\R = g^{\mu\nu}R_{\mu\nu} + v^{\alpha\beta}S_{\alpha\beta} = R+S \label{bundle ricci curvature}
	\end{equation}
	where
    \begin{equation}
         R_{\mu\nu} = R^\kappa_{\mu\nu\kappa} =  \delta_\kappa L^\kappa_{\mu\nu} - \delta_\nu L^\kappa_{\mu\kappa} + L^\rho_{\mu\nu}L^\kappa_{\rho\kappa} - L^\rho_{\mu\kappa}L^\kappa_{\rho\nu} + C^\kappa_{\mu\alpha}\Omega^\alpha_{\nu\kappa} \label{d-ricci 1}
    \end{equation}
	with
	\begin{equation}\label{Omega}
		\Omega^\alpha_{\nu\kappa} = \dder{N^\alpha_\nu}{x^\kappa} - \dder{N^\alpha_\kappa}{x^\nu}
	\end{equation}
    the non-holonomy coefficients. For a more detailed analysis, see \cite{Miron1994}.
	
	\subsection{Field equations}
	A Hilbert-like action on $TM$ is \cite{Konitopoulos:2021eav}
	\begin{equation}\label{Hilbert like action}
		K = \int_{\mathcal N} d^8\mathcal U \sqrt{|\Gd|}\, \R + 2\kappa \int_{\mathcal N} d^8\mathcal U \sqrt{|\Gd|}\,\mathcal L_M
	\end{equation}
	for some closed subspace of the Lorentz tangent bundle $\mathcal N\subset TM$, where $|\Gd|$ is the absolute value of the metric determinant, $\mathcal L_M$ is the Lagrangian of the matter fields, $\kappa$ is a constant and
	\begin{equation}
		d^8\mathcal U = \de x^0 \wedge \ldots \wedge\de x^3 \wedge \de y^4 \wedge \ldots \wedge \de y^7
	\end{equation}
	is the 8-volume element.

	Variation of the action with respect to $g_{\mu\nu}$, $v_{\alpha\beta}$ and $N^\alpha_\kappa$ yields the field equations:
	\begin{align}
		& \overline R_{\mu\nu} - \frac{1}{2}({R}+{S})\,{g_{\mu\nu}} + \left(\delta^{(\lambda}_\nu\delta^{\kappa)}_\mu - g^{\kappa\lambda}g_{\mu\nu} \right)\left(\mathcal D_\kappa\mathcal T^\beta_{\lambda\beta} - \mathcal T^\gamma_{\kappa\gamma}\mathcal T^\beta_{\lambda\beta}\right)  = \kappa {T_{\mu\nu}} \label{feq1}\\
		& S_{\alpha\beta} - \frac{1}{2}({R}+{S})\,{v_{\alpha\beta}} + \left(v^{\gamma\delta}v_{\alpha\beta} - \delta^{(\gamma}_\alpha\delta^{\delta)}_\beta \right)\left(\mathcal D_\gamma C^\mu_{\mu\delta} - C^\nu_{\nu\gamma}C^\mu_{\mu\delta} \right) = \kappa {Y_{\alpha\beta}} \label{feq2}\\
		& g^{\mu[\kappa}\partial_{\alpha}L^{\nu]}_{\mu\nu} +  2 \mathcal T^\beta_{\mu\beta}g^{\mu[\kappa}C^{\lambda]}_{\lambda\alpha} = \frac{\kappa}{2}\mathcal Z^\kappa_\alpha \label{feq3}
	\end{align}
	with
	\begin{align}
		T_{\mu\nu} &\equiv - \frac{2}{\sqrt{|\Gd|}}\frac{\Delta\left(\sqrt{|\Gd|}\,\mathcal{L}_M\right)}{\Delta g^{\mu\nu}} = - \frac{2}{\sqrt{-g}}\frac{\Delta\left(\sqrt{-g}\,\mathcal{L}_M\right)}{\Delta g^{\mu\nu}}\label{em1}\\
		Y_{\alpha\beta} &\equiv -\frac{2}{\sqrt{|\Gd|}}\frac{\Delta\left(\sqrt{|\Gd|}\,\mathcal{L}_M\right)}{\Delta v^{\alpha\beta}}  = -\frac{2}{\sqrt{-v}}\frac{\Delta\left(\sqrt{-v}\,\mathcal{L}_M\right)}{\Delta v^{\alpha\beta}}\label{em2}\\
		\mathcal Z^\kappa_\alpha &\equiv -\frac{2}{\sqrt{|\Gd|}}\frac{\Delta\left(\sqrt{|\Gd|}\,\mathcal{L}_M\right)}{\Delta N^\alpha_\kappa} = -2\frac{\Delta\mathcal{L}_M}{\Delta N^\alpha_\kappa}\label{em3}
	\end{align}
	where $\mathcal L_M$ is the Lagrangian of the matter fields, $\delta^\mu_\nu$ and $ \delta^\alpha_\beta$ are the Kronecker symbols, $|\Gd|$ is the absolute value of the determinant of the total metric \eqref{bundle metric}, with
    \begin{equation}\label{overline_ricci}
        \overline R_{\mu\nu} \equiv R_{\mu\nu} - C^\kappa_{\mu\alpha}\Omega^\alpha_{\nu\kappa}
    \end{equation}
    and
	\begin{equation}\label{torsion}
		\mathcal{T}_{\nu\beta}^{\alpha} = \partial_{\beta} N_{\nu}^{\alpha} - L_{\beta\nu}^{\alpha}
	\end{equation}
	are torsion components, where $L_{\beta\nu}^{\alpha}$ is given by \eqref{metric d-connection 2}. From \eqref{bundle metric} it immediatly follows that $\sqrt{|\Gd|} = \sqrt{g}\sqrt{v}$, with $g, v$ the determinants of the metrics $g_{\mu\nu}, v_{\alpha\beta}$ respectively. The constant $\kappa$ is found by taking the GR limit of the theory: 
	\begin{equation}
		\kappa = 8 \pi
	\end{equation}
	where we assume units where $c=1$ (the speed of light in vacuum) and $G=1$ (the Newtonian gravitational constant), a condition that holds for the rest of this work unless specified otherwise.
	
	\subsection{The Finsler-Randers metric}
	A special class of Finsler metric functions is the Finsler-Randers metric function, defined as:
	\begin{equation}\label{finsler_randers_metric_function}
		F(x,y) = \sqrt{g_{\mu\nu}(x) y^\mu y^\nu} + A_{(R)\gamma}(x) y^\gamma
	\end{equation}
	where $g_{\mu\nu}(x)$ is a pseudo-Riemannian Lorentzian metric tensor and $|A_{(R)\gamma}(x)| \ll 1$.
	The metric tensor derived from \eqref{finsler_randers_metric_function} by using the rule \eqref{finsler_metric} is a Finsler-Randers metric tensor.

	A notable application of this metric class is the Schwarzschild-Randers metric, which has been studied in a recent series of works \cite{Silagadze2010,Li2020,Kapsabelis2021,Ke-JianHe2024,Dehkordi2025,Yao2025}. The pseudo-Riemannian metric $g_{\mu\nu}(x)$ takes the form of the classic Schwarzschild metric:
	\begin{equation}
		ds^2 = \left( 1 - \frac{R_S}{r} \right) dt^2 - \left( 1 - \frac{R_S}{r} \right)^{-1} dr^2 - r^2 d\Omega^2
	\end{equation}
	where
	\begin{equation}\label{schwarzschild_radius}
		R_S = 2 M
	\end{equation}
	is the Schwarzschild radius, $M$ the mass equivalent of the Schwarzschild body and
	\begin{equation}
		d\Omega^2 = d \theta^2 + \sin^2\theta \, d \phi^2
	\end{equation}
	the metric of the 2-sphere.
	The nonlinear connection is:
	\begin{equation}
		N^\alpha_\mu = \frac{1}{2} y^\gamma g^{\alpha\beta}\partial_\mu g_{\beta\gamma}
	\end{equation}
	where the metric $g_{\mu\nu}$ is lift to the vertical subspace via the relation
	\begin{equation}
		g_{\alpha\beta} = \tilde \delta^\mu_\alpha \tilde \delta^\nu_\beta g_{\mu\nu}
	\end{equation}
	with $\tilde\delta^\mu_\alpha$ equal to $1$ if $\alpha = \mu + 4$ and equal to $0$ otherwise.

	The Randers one-form $A_{(R)\gamma}(x)$ is obtained as the solution of the field equations \eqref{feq1} - \eqref{feq3}, with the condition $|A_{(R)\gamma}(x)| \ll 1$ \cite{triantafyllopoulos2020}:
	\begin{equation}\label{Asolution}
		A_{(R)\gamma}(x) = \left[\tilde A_0 \left|1-\frac{R_S}{r} \right|^{1/2}, 0, 0, 0 \right]
	\end{equation}
	with $\tilde A_0$ a constant.

	\subsection{Reissner-Nordström and Kerr metrics}
	In this work, we will study Finsler-Randers generalizations of two classic GR solutions, the Reissner-Nordström metric:
	\begin{equation}\label{reissner-nordstrom}
		ds^2 = \left( 1 - \frac{R_S}{r} + \frac{Q^2}{r^2} \right) \, dt^2  - \left( 1 - \frac{R_S}{r} + \frac{Q^2}{r^2} \right)^{-1} \, dr^2  -  r^2 \, d\theta^2  - ~ r^2\sin^2\theta \, d\varphi^2
	\end{equation}
	and the Kerr metric:
	\begin{equation}\label{kerr}
		ds^2 = -\left(\frac{dr^2}{\Delta} + d\theta^2 \right) \rho^2 + \left(dt - a \sin^2 \theta \, d\phi \right)^2 \frac{\Delta}{\rho^2} - \left(\left(r^2 + a^2 \right) d\phi - a \, dt \right)^2 \, \frac{\sin^2 \theta}{\rho^2}
	\end{equation}
	where 
	\begin{equation}\label{kerr_delta}
		\Delta=r^2-R_Sr+a^2
	\end{equation}
	\begin{equation}\label{reduced_angular_momentum}
		a = \frac{J}{M}
	\end{equation}
    \begin{equation}\label{kerr_rho}
        \rho^2 = r^2 + a^2 \cos^2\theta
    \end{equation}
	$Q$ is the charge of the body, $J$ the angular momentum of the body and $R_S$ is the Schwarzschild radius.

\section{Finsler Reissner-Nordström spacetime}
\label{sec3}

The Reissner-Nordström (RN) metric \eqref{reissner-nordstrom} describes a spherically symmetric spacetime, where $f(r) = g_{tt} = 1 - \frac{2M}{r} + \frac{Q^2}{r^2}$. The solution of the field equations \eqref{feq1}-\eqref{feq3} for the Randers one-form is therefore inferred from the general solution of the spherically symmetric problem \cite{triantafyllopoulos2020} to be
\begin{equation}\label{Asolution_RN}
    A_{(R)\gamma}(x) = \left[\tilde A_0 \left|1-\frac{2M}{r} + \frac{Q^2}{r^2} \right|^{1/2}, 0, 0, 0 \right]
\end{equation}
where $\tilde A_0$ is an unspecified normalisation constant that satisfies the perturbative assumption $|A_{(R)\gamma}(x)| \ll 1$.

The line element of a curve in a Randers-type spacetime is defined to be 
\begin{equation}
    ds = \sqrt{g_{\mu\nu} dx^\mu dx^\nu} + {A_{(R)\mu}} dx^\mu
    \label{eq56}
\end{equation}
with ${A_{(R)\mu}}$ being the anisotropy-inducing Randers one-form, and the Lagrangian is given as $\mathcal{L} = \frac{ds}{d\lambda}$ with respect to an affine parameter $\lambda$ such that $\sqrt{g_{\mu\nu} \frac{dx^\mu}{d\lambda} \frac{dx^\nu}{d\lambda}} = \sigma$ is a constant along the curve. We posit that physical trajectories are geodesics of our spacetime which extremise the length $\int_A ^B ds$ between points $A$ and $B$ of the manifold, and are therefore found by solving the corresponding Euler-Lagrange equations. We characterise, however, these geodesics as timelike, lightlike or spacelike depending on the sign of the Riemannian line element:
\begin{itemize}
    \item Timelike: $\sigma^2 = 1$ along the curve,
    \item Lightlike: $\sigma^2 = 0$ along the curve, and
    \item Spacelike: $\sigma^2 = -1$ along the curve.
\end{itemize}
This choice is made in order to preserve the invariance of the speed of lightlike signals for all observers in our spacetime. Because the spacial metric can always be made locally Minkowski by a diagonalising transformation of the spacial coordinates, in this locally Lorentz frame $(t,X)$ the Riemannian line element can be written as $\sqrt{g_{\mu\nu} dx^\mu dx^\nu} = \sqrt{dt^2 - dX^2}$. This choice of reference frame allows us to see that setting $\sigma = 0$ amounts to $\big|\frac{dX}{dt}\big| = 1$ for a lightlike trajectory. 

The Euler-Lagrange equations for $\mathcal{L}$ of a unit charge particle, with the addition of an electromagnetic tensor $F_{\mu\nu}$, furnish the geodesic equations
\begin{equation}
    \ddot{x}^\mu + {\Gamma^\mu }_{\nu\rho} \dot{x}^\nu \dot{x}^\rho - \sigma g^{\mu\nu}{\Phi}_{(R)\nu\rho} \dot{x}^{\rho} = \sigma g^{\mu\nu}F_{\nu\rho} \dot{x}^{\rho}
\end{equation}
which we will write compactly by defining $\Phi_{\mu\nu} = F_{\mu\nu} + {\Phi_{(R)\mu\nu}}$, or $\Phi_{\mu\nu} = \partial_\mu A_\nu - \partial_\nu A_\mu$ with $A_\mu ={A_{(R)\mu}} + {A_{(EM)\mu}}$ where ${A_{(EM)\mu}}$ is the electromagnetic 4-potential:
\begin{equation}
    \ddot{x}^\mu + {\Gamma^\mu }_{\nu\rho} \dot{x}^\nu \dot{x}^\rho = \sigma g^{\mu\nu}\Phi_{\nu\rho} \dot{x}^{\rho}.
\end{equation}
We take the indices to correspond to the usual spherical coordinates as $(x^0,x^1,x^2,x^3) \equiv (t,r,\theta,\phi)$. The unified one-form $A_\mu$ is only nonzero in its time component, which we denote as $A_0 = F(r)$:
\begin{equation}
    F(r) = {A_{(EM)0}} + {A_{(R)0}}  =  \frac{Q}{r} + \tilde{A}_0 \sqrt{f(r)}
\end{equation}
where, for the Reissner-Nordström metric, $f(r) = 1 - \frac{2M}{r} + \frac{Q^2}{r^2}$.

The resulting equations for $\mu=2,3$ are readily integrated to give, respectively, 
\begin{equation}
\begin{gathered}
    \frac{d}{d\lambda} (r^2 \dot\theta) = -2 r^2 \cot\theta \hspace{1mm} \dot\theta \dot\phi \\
    \frac{d}{d\lambda} (r^2 \dot\phi) = \frac{1}{2} r^2 \sin 2\theta (\dot\phi)^2
\end{gathered}
\end{equation}
which we shall simplify by restricting our attention to trajectories that lie on the equatorial plane $\theta = \frac{\pi}{2}$ with no loss of generality, as spherical symmetry dictates that all orbits be confined in the plane spanned by the radial direction vector and the velocity vector. Of the above equations one is therefore satisfied trivially and the other yields the angular momentum conservation law
\begin{equation}
    r^2 \dot\phi = l
\end{equation}
where $l$ is an integral of motion corresponding to the orbital angular momentum.

The $\mu = 0$ equation reads
\begin{equation}
    \ddot{t} + \frac{f'(r)}{f(r)} \dot{t}\dot{r} = -\sigma \frac{F'(r)}{f(r)} \dot{r} 
\end{equation}
which, by recalling that $\frac{d}{d\lambda} F(r) = F'(r) \dot{r}$, gives us the energy integral 
\begin{equation}
    f(r) \dot{t} = -\sigma F(r) + \mathcal{E}
\end{equation}
where $\mathcal{E}$ is the resulting integral of motion. Note that $f'(r)$ denotes differentiation of $f(r)$ with respect to the variable $r$.

Lastly we write the $\mu = 1$ equation
\begin{equation}
    \ddot{r} - \frac{f'(r)}{2 f(r)} \left(\dot{r}\right)^2 + \frac{f(r) f'(r)}{2} \left(\dot{t}\right)^2 - r f(r) \left(\sin^2 \theta \left(\dot{\phi}\right)^2 + \left(\dot{\theta}\right)^2 \right) = -\sigma f(r) F'(r) \dot{t} .
\end{equation}
Substituting the previous expressions for $\dot{t}$ and $\dot{\phi}$ in terms of $\mathcal{E}$ and $l$ and rearranging permits us to integrate the equation, introducing an integration constant $\epsilon$, reaching the resulting expression
\begin{equation}
    (\dot{r})^2 + f(r) \left(\frac{l^2}{r^2} + \epsilon\right) = \mathcal{E}^2 + \sigma ^2F^2(r) - 2 \sigma \mathcal{E} F(r)
\end{equation}
which completes the set of equations required to describe a geodesic trajectory in our Finsler-like spacetime, as at every point all $y^\alpha$ coordinates are given by the tangent vector components $y^\alpha (\lambda) = \tilde{\delta}^\alpha _\mu \frac{dx^\mu(\lambda)}{d\lambda}$.

We can immediately identify the integration constant $\epsilon$ as the square of the Riemannian interval, by evaluating the expression $\sigma^2 = g_{\mu\nu} \dot x ^\mu \dot x ^\nu$, which gives us the equation
\begin{equation}
     (\dot{r})^2 + f(r) \left(\frac{l^2}{r^2} + \sigma^2 \right) = \mathcal{E}^2 + \sigma ^2F^2(r) - 2 \sigma \mathcal{E} F(r)
\end{equation}
and so we conclude that $\epsilon = \sigma^2$. One can therefore proceed to restrict these geodesic equations to lightlike or timelike trajectories by setting $\sigma = 0$ or $\sigma = 1$ respectively. In the former case we observe that all of our equations return to their GR counterparts as all of the additional, perturbing terms are weighted by $\sigma$ and are therefore lost. That is, lightlike curves satisfy
\begin{equation}
\begin{gathered}
    f(r) \dot{t} = \mathcal{E} \\
    (\dot{r})^2 + f(r) \frac{l^2}{r^2} = \mathcal{E}^2 ,
\end{gathered}
\end{equation}
which are the same equations that define lightlike trajectories in a classical Reissner-Nordström spacetime. We therefore find no observable imprint of the Finsler-Randers geometry on lightlike trajectories, and expect all related quantities, such as the amount of the deflection of light rays, to remain unchanged. 

On the other hand, timelike trajectories   satisfy
\begin{equation}
\begin{gathered}
    f(r) \dot{t} = - F(r) + \mathcal{E} \\
    (\dot{r})^2 + f(r) \left(\frac{l^2}{r^2} + 1 \right) = \mathcal{E}^2 + F^2(r) - 2 \mathcal{E} F(r)
\end{gathered}
\end{equation}
where we identify the effective radial potential 
\begin{equation}
    V_{eff}(r) = \frac{f(r)}{2} \left(\frac{l^2}{r^2} + 1 \right) - \frac{1}{2}F^2(r) + \mathcal{E} F(r)
\end{equation}
and observe a correction term in the energy of the orbit in addition to the standard effective potential of a spherically symmetric spacetime
\begin{equation}
    V_{eff, GR}(r) = \frac{f(r)}{2} \left(\frac{l^2}{r^2} + 1 \right).
\end{equation}

\begin{figure}[h!]
\centering
\includegraphics[width=0.6\linewidth]{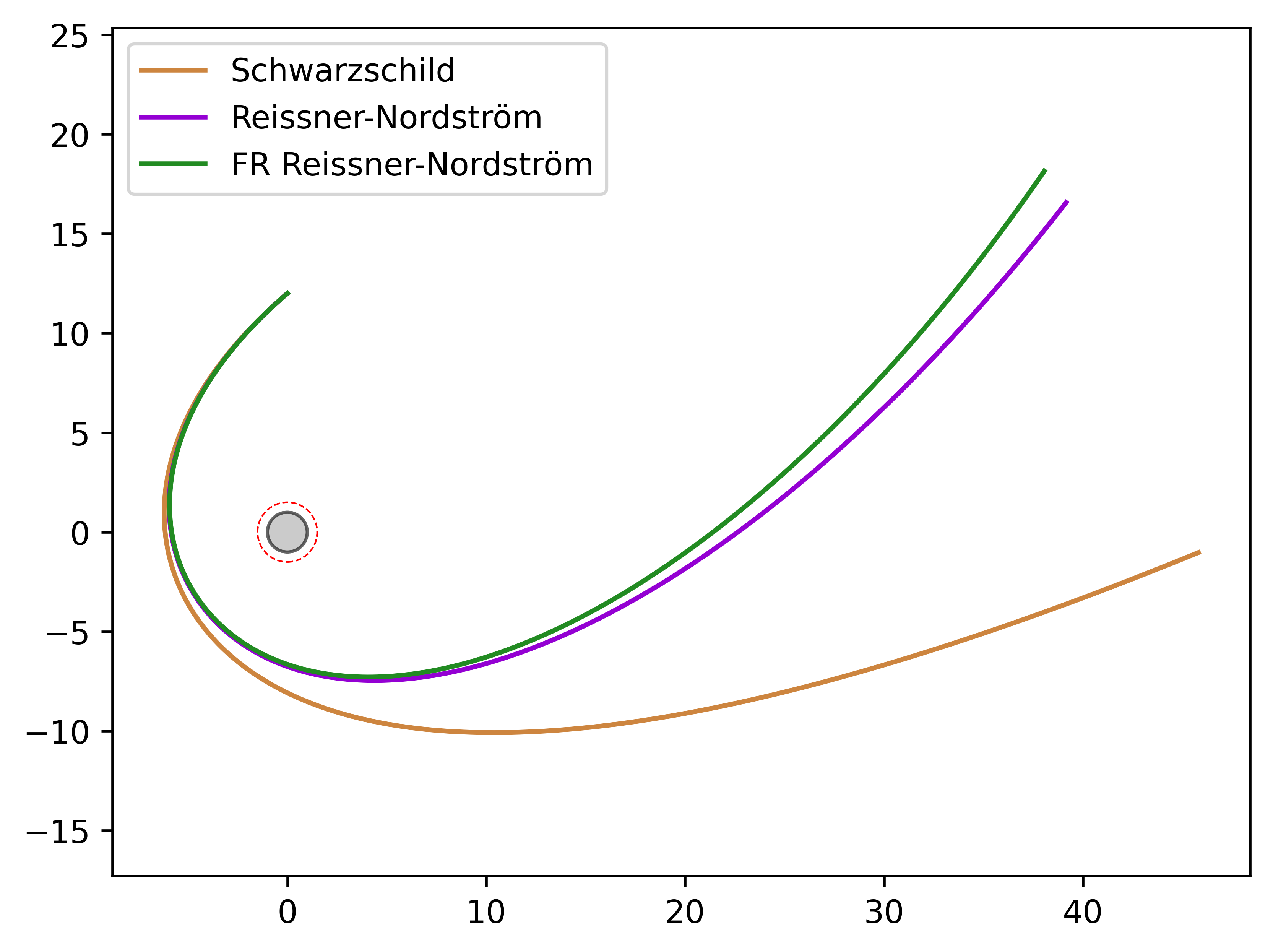} 
\includegraphics[width=0.6\linewidth]{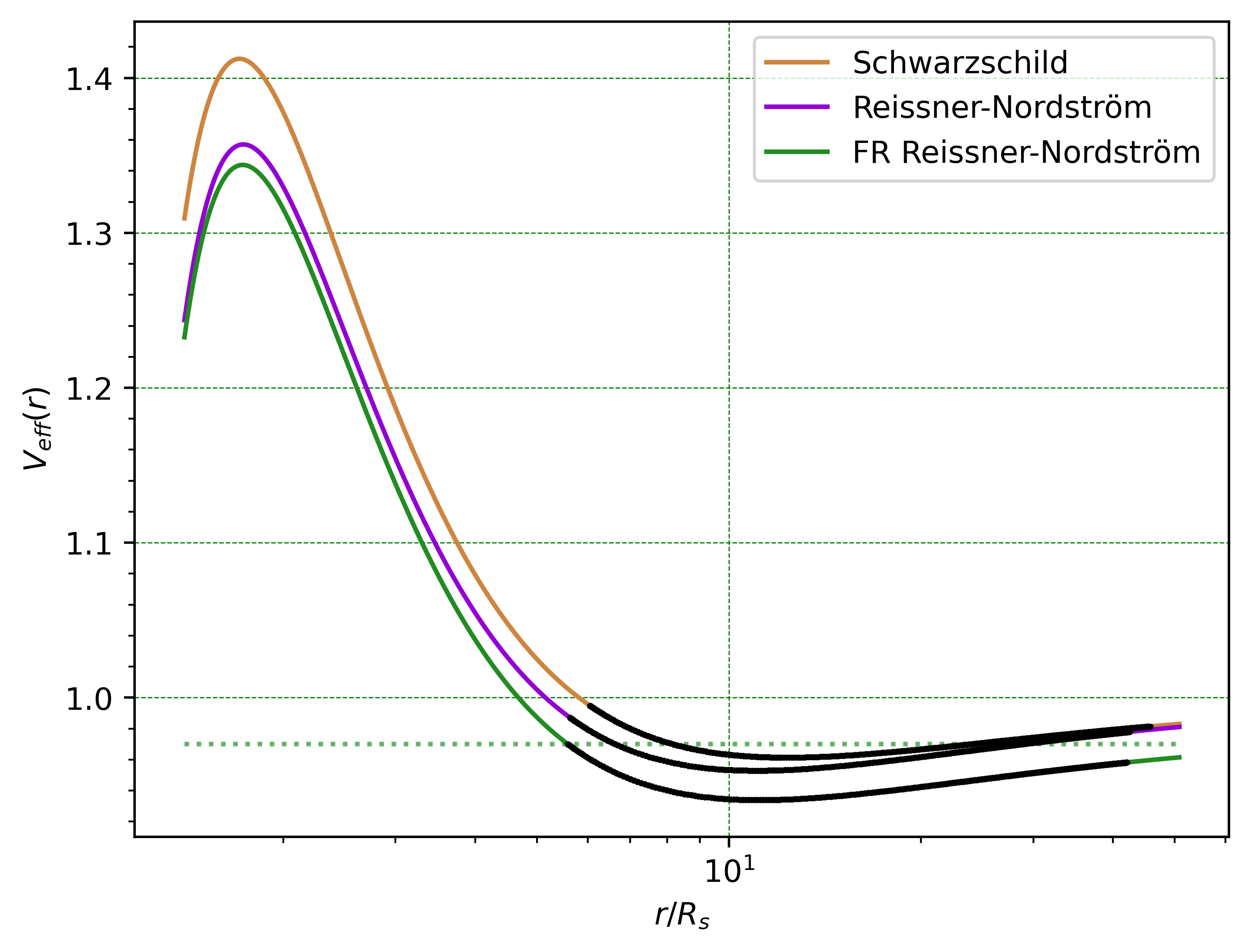}
\captionsetup{width=0.85\linewidth}
\caption{(up) Bounded timelike orbits of a unit mass and charge particle in a Randers Reissner-Nordström, classical Reissner-Nordström and classical Schwarzschild spacetime. The black hole Schwarzschild radius $R_s$ is set to 1 and the photonsphere (red dashed circle), identical for all three spacetimes, lies at $1.5 R_s$. 
(down) The radial motion of the orbits (black trail) on the respective effective potential. 
For all orbits we choose $l=2.65$ and determine the initial conditions so that $E = 0.97$ for the FRN orbit (green dashed line). For the spacetime parameters we have additionally set $Q = 0.1M$ and $\tilde A_0 = 0.01$. The radial motion integration is performed using a nested 4th order Runge-Kutta algorithm.}
\label{fig:veffs}
\end{figure}

In Figure \ref{fig:veffs} we qualitatively plot a timelike orbit of a unit mass and charge particle in the above potential. The motion is compared to the respective orbits in the standard GR Schwarzschild and Reissner-Nordström spacetimes with identical initial conditions, whereas the orbital energy differs as a result of the modified energy integral. The orbits are quickly precessing Keplerian ellipses and fall back to the classical Schwarzschild orbit in the zero charge and $\tilde A_0$ limit.

\begin{figure}[ht]
\centering
\includegraphics[width=0.7\linewidth]{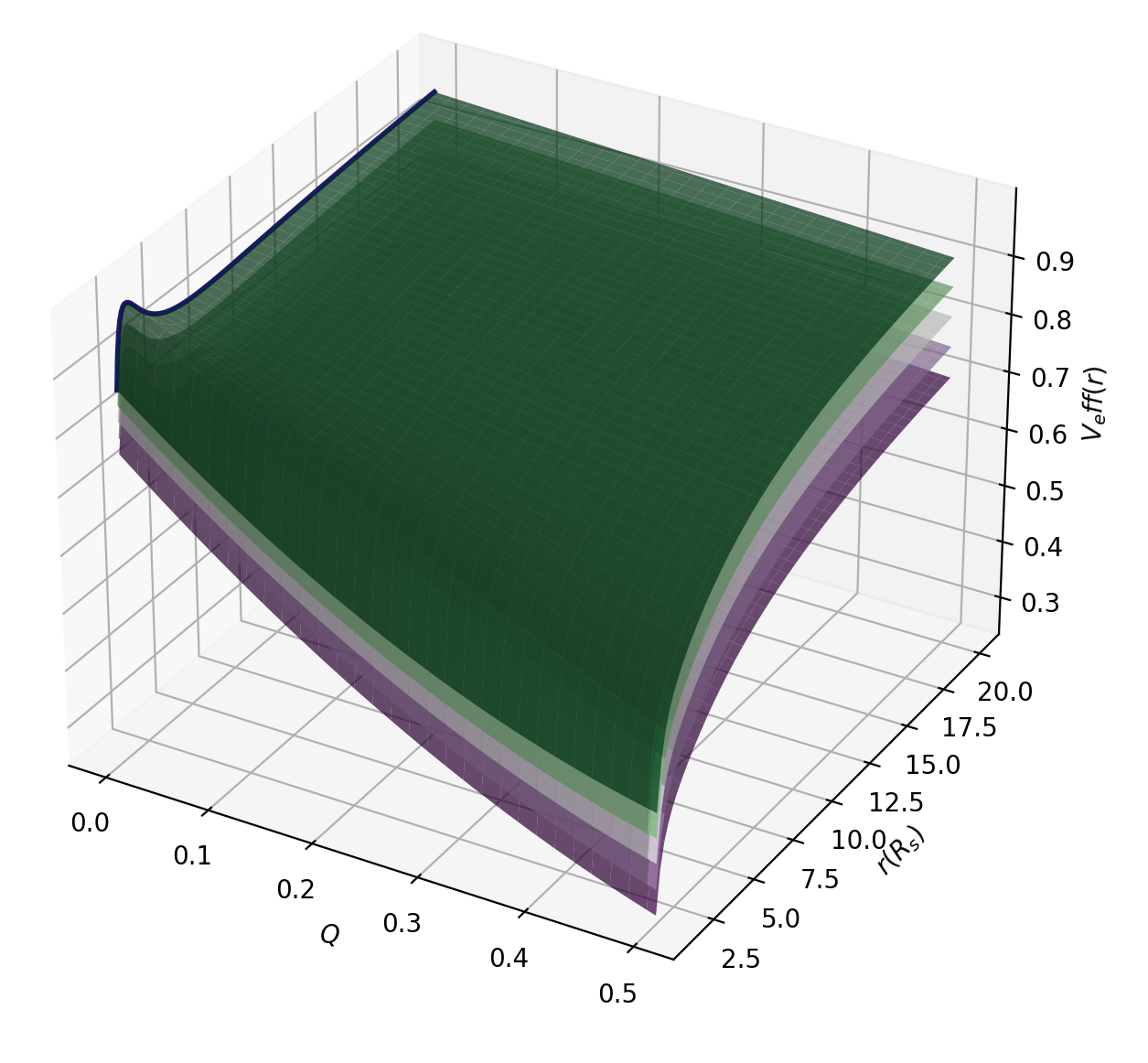}
\captionsetup{width=0.85\linewidth}
\caption{The radial effective potential for timelike trajectories in the Finsler-Randers Reissner-Nordström spacetime for a massive, charged particle. The potential is taken as a function of the black hole charge $Q$ and plotted for values of the Randers one-form magnitude $\tilde A_0$ ranging linearly from 0 (green) to $0.1$ (purple). It is defined with $(\dot r)^2 + V_{eff}(r) = E^2$ with parameters $l=2$, $E^2 = 0.97$ in units where $c=1, G=1$ and $M = 0.5$. The bold line on the $\tilde A_0 = 0$ sheet, $Q=0$, represents the Schwarzschild effective potential.}
\label{fig:veff_func}
\end{figure}

In Figure \ref{fig:veff_func} we illustrate the dependence of the effective potential on the black hole charge $Q$ and the Randers one-form magnitude $\tilde A_0$, stressing that our solution is found to the first order in $\tilde A_0$. The potential well that confines stable orbits is seen to descend and diminish in depth before it disappears, as $Q$ increases.

\section{A Kerr-like approach for a Finsler-Raders spacetime}
We will now treat the case of the Kerr spacetime with a Finsler-Randers perturbation to the line element. The solution of the field equations is derived analytically for the $A_5, A_6$ components of the Randers one-form, which are independently found to vanish, and numerically for the $A_4,A_7$ components which satisfy a nested PDE system. We find that any nonzero components of $A_\mu$ diverge at the limit of radial infinity, and go on to restrict our region of interest in the vicinity of the black hole so that our assumption that $A_\mu$ is of perturbative nature and negligible in terms of order that is quadratic or higher is respected. 

The following should therefore be understood as an application of a first-order scheme that is only locally applicable as, in greater distances, the full nonperturbative solution must be utilised. Depending on the latter's behaviour, one may find either that the divergence is no more present, or that the complete solution for the Finsler-Randers (FR) Kerr spacetime diverges at infinity, in which case the model would exhibit some exotic or unphysical properties. One such example would be the influence of the black hole on geodesics on the asymptotically flat background far away from the black hole, diverting them from being straight lines.

\subsection{Solution of the field equations}

We proceed to employ the Kerr metric~\eqref{kerr} as an ansatz for the Riemannian part of the Finsler-Randers metric function~\eqref{finsler_randers_metric_function}. The Kerr metric describes a rotating, non-charged black hole in vacuum; thus, the momentum tensors vanish:
\begin{equation}
    T_{\mu\nu} = Y_{\alpha\beta} = 0.
\end{equation}
Moreover, similarly to the Reissner–Nordström case and  the Schwarzschild case studied in~\cite{triantafyllopoulos2020}, the Riemannian metric $g_{\mu \nu}$ is independent of $y$, implying that the Cartan tensor $C^\kappa_{\mu\alpha}$ vanishes. As a consequence, the Ricci tensor $\Bar{R}_{\mu \nu}$ reduces to its general relativity limit through~\eqref{overline_ricci}—which vanishes for the Riemannian Kerr metric—and the $S$-curvature tensor vanishes as well. Thus, equation~\eqref{feq1} reduces to the same form as~\eqref{reducedfe} for the RN case.

Assuming $|A_{(R)\gamma}| \ll 1$ and neglecting quadratic terms in $A_{(R)\gamma}$, following the same analysis as presented in~\cite{triantafyllopoulos2020} we arrive at
\begin{equation}
\begin{aligned}
&\partial_\mu \partial_\nu A_\gamma 
- \frac{1}{2} g^{\beta\delta} \partial_\nu g_{\delta\gamma} \, \partial_\mu A_\beta 
- \frac{1}{2} g^{\beta\delta} \partial_\mu g_{\delta\gamma} \, \partial_\nu A_\beta \\
&\quad + \frac{1}{4} A_\beta \Bigg(
\frac{1}{2} g^{\beta\epsilon} g^{\delta\zeta} \partial_\nu g_{\epsilon\delta} \, \partial_\mu g_{\gamma\zeta}
+ \frac{1}{2} g^{\beta\epsilon} g^{\delta\zeta} \partial_\mu g_{\epsilon\delta} \, \partial_\nu g_{\gamma\zeta} \\
&\quad - \partial_\mu g^{\beta\delta} \, \partial_\nu g_{\gamma\delta}
- \partial_\nu g^{\beta\delta} \, \partial_\mu g_{\gamma\delta}
- 2 g^{\beta\delta} \partial_\mu \partial_\nu g_{\gamma\delta}
\Bigg) \\
&\quad - \frac{1}{2} g^{\kappa\lambda} (\partial_\mu g_{\kappa\nu} + \partial_\nu g_{\kappa\mu} - \partial_\kappa g_{\mu\nu}) \left( \partial_\lambda A_\gamma - \frac{1}{2} A_\beta g^{\beta\delta} \partial_\lambda g_{\gamma\delta} \right) = 0 .
\end{aligned}
\label{eq72}
\end{equation}
We now proceed to solve \eqref{eq72} in order to determine the components of \( A_{(R)\gamma} \). Selecting \((\mu,\nu) = (0,0)\), \((0,3)\) and \(\gamma = 6\), we obtain the equations
\begin{equation}
  \begin{gathered}
   \partial_1 A_6 - \frac{A_6}{2} g^{22} \partial_1 g_{22} = 0 \\ 
   \partial_2 A_6 - \frac{A_6}{2} g^{22} \partial_2 g_{22} = 0,
  \end{gathered}
\end{equation}
which admit the solution \( A_6 = \tilde{A}_6 \rho \), where \( \tilde{A}_6 \) is a constant. Similarly, for \(\gamma = 5\), we obtain
\begin{equation}
  \begin{gathered}
   \partial_1 A_5 - \frac{A_5}{2} g^{11} \partial_1 g_{11} = 0 \\ 
   \partial_2 A_5 - \frac{A_5}{2} g^{11} \partial_2 g_{11} = 0,
  \end{gathered}
\end{equation}
which admit the solution \( A_5 = \tilde{A}_5 \frac{\rho}{\sqrt{\Delta}} \).  
In both the \(\gamma = 5\) and the \(\gamma = 6\) case, by substitution of the aforementioned solutions into the equation corresponding to \((\mu,\nu) = (1,1)\), one shows that it is satisfied only if \(\tilde{A}_5 =0\) and \( \tilde{A}_6= 0\). Thus, the components \( A_5\) and \( A_6 \) reduce to the trivial solutions
\begin{equation}
\label{eq:a5a6}
    A_5 = 0, \quad A_6 = 0.
\end{equation}

\begin{figure}[h]
\centering
\includegraphics[width=0.8\linewidth]{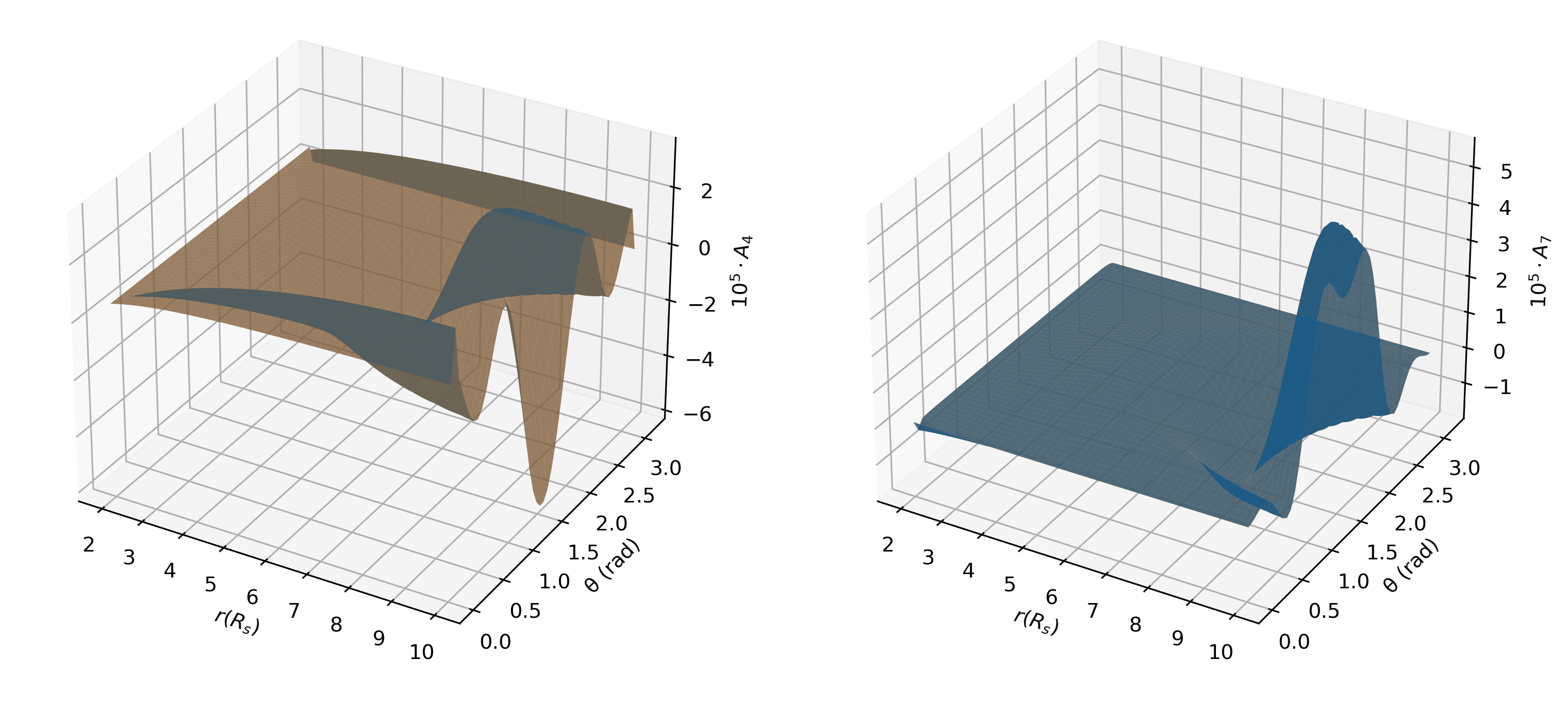}
\captionsetup{width=0.85\linewidth}
\caption{Numerical solutions for the $A_4$, $A_7$ components of the Randers one-form in the Kerr-Randers spacetime. The normalisation is arbitrary and was chosen for the initial grid point so that the maximum absolute values are in the order of $10^{-5}$. Beyond the radial cut-off the $A_4, A_7$ functions continue to increase indefinitely and diverge at radial infinity.}
\label{fig:a4_a7}
\end{figure}

Equations \eqref{eq72} for the components $A4$ and $A_7$ yield with the choices $(\mu,\nu) = (0,0), (0,3), (3,0), (3,3)$ the following nested set of PDEs:
\begin{equation}
\begin{gathered}
    \partial_1 A_7 - \frac{1}{2} (g^{00} \partial_1 g_{03} + g^{03} \partial_1 g_{33}) A_4 - \frac{1}{2} (g^{03} \partial_1 g_{03} + g^{33} \partial_1 g_{33}) A_7 = 0 \\
    \partial_2 A_7 - \frac{1}{2} (g^{00} \partial_2 g_{03} + g^{03} \partial_2 g_{33}) A_4 - \frac{1}{2} (g^{03} \partial_2 g_{03} + g^{33} \partial_2 g_{33}) A_7 = 0 \\
    \partial_1 A_4 - \frac{1}{2} (g^{00} \partial_1 g_{00} + g^{03} \partial_1 g_{03}) A_4 - \frac{1}{2} (g^{03} \partial_1 g_{00} + g^{33} \partial_1 g_{03}) A_7 = 0 \\
    \partial_2 A_4 - \frac{1}{2} (g^{00} \partial_2 g_{00} + g^{03} \partial_2 g_{03}) A_4 - \frac{1}{2} (g^{03} \partial_2 g_{00} + g^{33} \partial_2 g_{03}) A_7 = 0 .
\end{gathered}
\end{equation}
The solutions are determined numerically up to a normalising factor and depicted in Figure \ref{fig:a4_a7}. The nested set of PDEs was integrated within a confined region of the $(r,\theta)$ plane in the ranges $r/R_s \in (2, 10)$ and $\theta \in (0, \pi)$ by evaluation of the partial derivative on grid points and propagation the function in the respective direction through a simple Euler-like algorithm. The data points are then fit with a Fourier series of the form $f_{fit} (r,\theta) = \sum_{n,m} ( \sin (\frac{n \pi r}{8}) + \cos( \frac{n \pi r}{8}))( \sin (\frac{m \theta}{2}) + \cos( \frac{m \theta}{2}))$ up to terms of $n_{max}=15,m_{max}=15$ and so the solution is obtained within the restricted region in terms of its Fourier coefficients. 

Both functions appear to diverge in an oscillating manner at the limit of infinite $r$ and so the solutions are restricted to a finite region with a radial cutoff such that the first order approximation in the magnitude of $A_\mu$ is respected, and the maxima of the absolute values of its components remain well below 1. One should note that these solutions have not been verified to satisfy all choices of indices $\mu,\nu$ in \eqref{eq72} as no complete explicit expressions have been obtained.

\subsection{Geodesics in the Kerr FR spacetime}

The procedure for obtaining the geodesics follows the same principles that were stated in Section \ref{sec3} for the Reissner-Nordström spacetime. Minimisation of the action given by the Kerr line element with a Randers perturbation, as shown in \eqref{eq56}, yields the vacuum equations of motion 
\begin{equation}
    \ddot x^\mu + \Gamma^\mu _{\nu\rho} \dot x^\nu \dot x^\rho = \sigma g^{\mu\nu} \Phi_{(R)\mu\nu}
\end{equation}
with $\Phi_{(R)\mu\nu} = \partial_\mu A_{(R)\nu} - \partial_\nu A_{(R)\mu}$. As before we note that lightlike geodesics, found with $\sigma = 0$, are identical to their Riemannian counterparts, and so all associated quantities remain invariant in transitioning to the Finsler-Randers framework.

\begin{figure}[h!]
\centering
\includegraphics[width=0.7\linewidth]{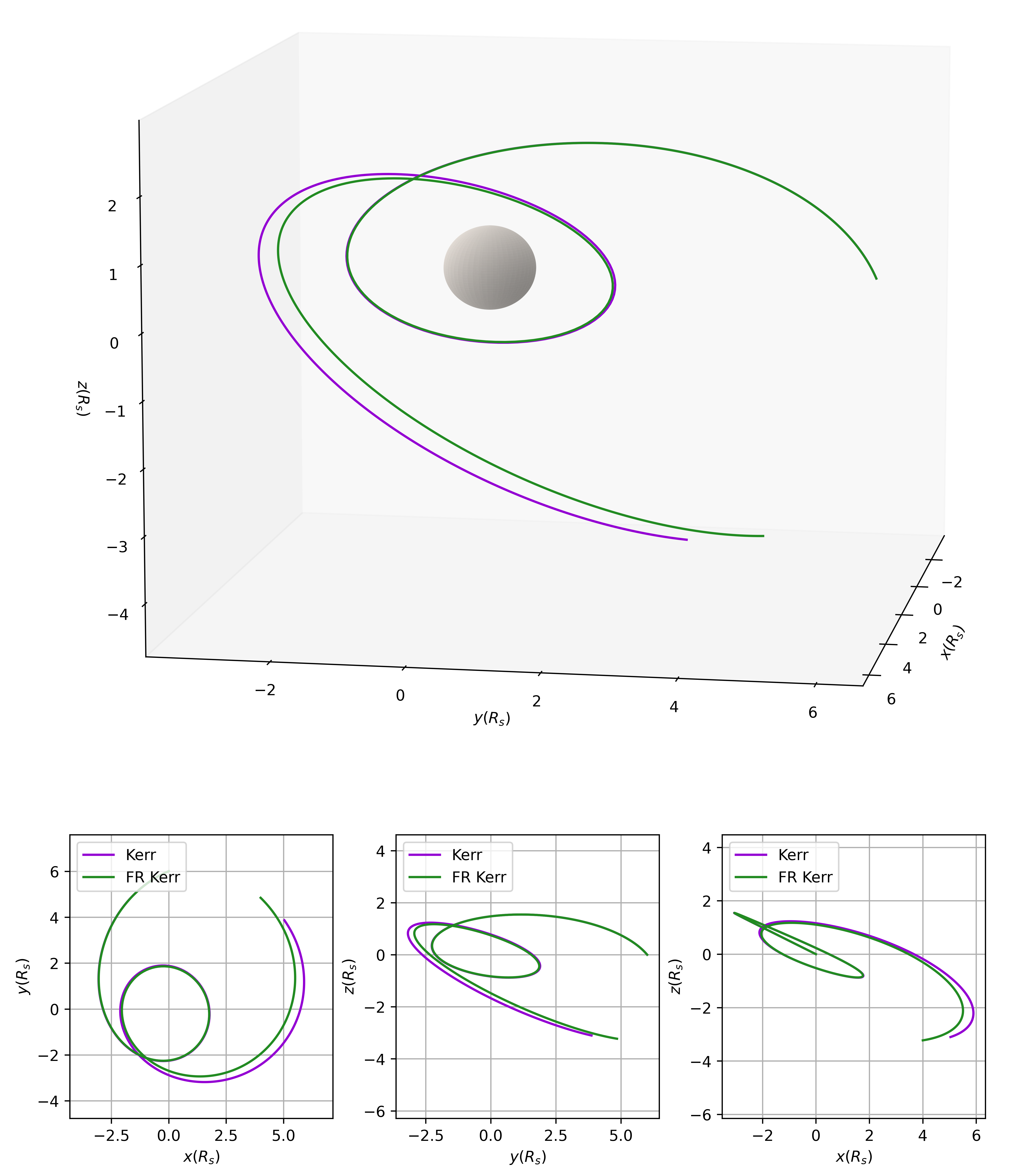}
\captionsetup{width=0.85\linewidth}
\caption{Orbit of a timelike particle in the Kerr and Kerr-Randers (FR) spacetime plotted in Boyer-Lindquist coordinates. The particle is given identical initial conditions and the equations of motion are then integrated numerically, in the latter case with the inclusion of the perturbing Randers one-form components $A_4, A_7$. The initial conditions are chosen such that the orbits are restricted within the $(r,\theta)$ domain where the solutions of the $A_4,A_7$ components are defined. The black hole's parameters have been set to $M = 0.5$ and $\alpha = 0.3$.}
\label{fig:Kerr_orbits}
\end{figure}

Timelike geodesics are characterised by $\sigma = 1$ and therefore differ from the Riemannian trajectories. We illustrate this difference by performing a numerical integration of one such orbit in Figure \ref{fig:Kerr_orbits}, where the timelike motion in the FR Kerr spacetime is contrasted with the corresponding geodesic path in the classical Kerr spacetime obtained via the same initial conditions. The integration is done using a 4-dimensional nested Runge-Kutta 4 algorithm on the Boyer-Lindquist coordinate basis, and graphed using an asymptotically Cartesian grid $(\rho \cos \phi, \rho \sin \phi, z)$ where $\rho = \sqrt{r^2 + \alpha^2} \sin \theta, z = r \cos \theta$. Since we do not have an analytical expression for $A_4, A_7$ but rather a Fourier series approximation, we integrate the geodesic equation $\ddot x^\mu + \Gamma^\mu _{\nu\rho} \dot x^\nu \dot x^\rho = \sigma g^{\mu\nu} \Phi_{(R)\nu\rho} \dot x^\rho$ directly with numerical methods.

\subsection{Quasinormal modes of background fields}

A generic field $\phi^I (x,y)$ with a collection of tensor or spinor indices $I$ will be governed by a Lagrangian density $\mathcal{L} [\phi^I, \partial_\mu \phi^I]$ that furnishes the field equations upon minimisation of the action with respect to $\phi^I$. Transferring this field to our 8-dimensional Finslerian manifold, we may generalise the Lagrangian $\mathcal{L}_F [\phi^I, \partial_\mu \phi^I, \partial_\alpha \phi^I]$ to admit derivatives of the field along the tangent directions as well.

We may postulate, however, that the field $\phi^I$ be constant over the vertical part of the tangent bundle, that is, that the field is uniquely determined, and singled valued, everywhere in spacetime. This has been utilised in the Appendix to justify that the field's energy-momentum tensor coincides with its Riemannian energy-momentum tensor. With this ansatz all derivatives along the tangent directions thus vanish and the Lagrangian reduces to its initial form $\mathcal{L}_R [\phi^I, \partial_\mu \phi^I]$. Variation of the action with respect to the field $\phi^I$ will therefore produce the same equations of motion one would get in the Riemannian setting, supplemented by the fact that all inner products along spacial (that is, non-tangent) vector, including derivatives, are taken with the spacial metric $g_{\mu\nu}$ and hence no additional contribution from our Randers perturbation emerges, the field equations consequently remaining the same as those used in the Riemannian setting, and admitting the same oscillatory modes.  

We will demonstrate the above for a massive Klein-Gordon scalar field $\phi(x,y)$. We write a Klein-Gordon Lagrangian density with the addition of a quadratic term on the tangential space
\begin{equation}
    \frac{1}{\sqrt{-g}}\mathcal{L} = \frac{1}{2} g^{\mu\nu} D_\mu \phi \cdot D_\nu \phi + \frac{1}{2} v^{\alpha\beta} D_\alpha \phi \cdot D_\beta \phi + \frac{1}{2} m^2 \phi^2 .
\end{equation}
The Euler-Lagrange equations with respect to $\phi$ give us the field equation
\begin{equation}
    g^{\mu\nu} D_\mu D_\nu \phi + v^{\alpha\beta} D_\alpha D_\beta \phi= m^2 \phi
\end{equation}
which can be examined for oscillatory modes in the entire 8-dimensional manifold. Choosing to restrict the dependence of $\phi$ to the horizontal part results in the standard massive Klein-Gordon equation
\begin{equation}
    g^{\mu\nu} D_\mu D_\nu \phi = m^2 \phi
\end{equation}
that one obtains classically. From this point onward the procedure of solving for the field's quasinormal modes reduces to the analysis performed by \cite{Konoplya:2006br} for a massive scalar field in the Kerr manifold.

\section{Conclusions}
In this article, we extended the Finsler–Randers cosmological framework to incorporate charged and rotating spacetimes, focusing on generalizations of the Reissner–Nordström and Kerr solutions. Building upon previous investigations of spherically symmetric configurations, our analysis explored the consequences of local anisotropy introduced via internal coordinates on the structure of black hole solutions.

Through the modified gravitational field equations we derived  solutions for the perturbing Randers one-form on the manifold and analyzed the geodesic motion of the test particles. Within our formulation of timelike and lightlike character based on the Riemannian line element, these results confirm that light-like geodesics remain unaffected by Finsler–Randers perturbations, preserving observational predictions from General Relativity (GR). However, time-like geodesics exhibit significant deviations, particularly in the form of modified energy integrals and effective potentials. These corrections manifest as shifts in orbital dynamics, including precession and energy variations, which vanish in the isotropic or uncharged limits.

The exploration of timelike orbits and the radial effective potential reveals key modifications in particle dynamics within the Finsler-Randers-Reissner-Nordström (FRRN) spacetime. We have qualitatively compared bounded orbits in the Finsler-Randers, standard Reissner-Nordström, and Schwarzschild spacetimes, where in the limit of vanishing charge and of vanishing perturbing field, the orbits approach the Schwarzschild solution (See figure \ref{fig:veffs}). We have examined the radial effective potential dependence on the perturbing one-form's magnitude $\tilde A_0$ and the black hole charge $Q$, with the stable region seen to become shallower and reduce the stability of orbits, eventually vanishing (See figure \ref{fig:veff_func}). These results highlight the significant influence of anisotropic perturbations on the spacetime structure and provide a framework for further investigations of deviations from General Relativity in locally anisotropic spacetime.

To further research the impact of Finslerian geometry on a rotating spacetime, we partially derived an analytical solution for the Randers field in the Kerr spacetime and supplemented our result with numerical solutions for the remaining components. By performing simulations of particle motion in the perturbed Kerr spacetime, we found similar deviations from the classical trajectories when it comes to timelike orbits. 

Our results demonstrate that the inclusion of internal anisotropic degrees of freedom leads to measurable deviations in the geometry, particularly near strong gravitational regions such as black hole horizons. Modifications in the causal structure indicate that Finsler–Randers gravity may offer an enriched phenomenology compared to its Riemannian counterpart.

This investigation highlights the potential of Finsler-Randers cosmology to serve as a viable extension of general relativity, capable of accommodating anisotropies and direction-dependent phenomena with a consistent geometric framework in FRRN and Kerr-like solutions. Further investigations may elucidate observational signatures of such deviations, particularly in high precision astrophysical or cosmological settings. Such systems allow one to study in particular the effect of modified timelike geodesics on the motion of particles in the vicinity of a black hole and the resulting emission spectra \cite{Rarras2024,rarras2024} which can, in principle, be contrasted to observations.

\acknowledgments{The authors would like to thank Prof. Garry Gibbons for valuable discussions and insightful comments during the early stages of this work.}

\appendix
In this Appendix we derive the solution for the anisotropy inducing Randers one-form $A_{(R)\mu\nu}$ which we will here denote as simply $A_{\mu\nu}$, using the generalised Einstein's equations as stated in \eqref{feq1}-\eqref{feq3}. The first of those, taking $S=0$ as one finds by taking the trace of the second equation, reads
\begin{equation}
    R_{\mu\nu} -  \frac{1}{2}R g_{\mu\nu} + (\delta^{(\lambda}_\nu \delta^{\kappa )}_\mu - g^{\kappa\lambda}g_{\mu\nu})(\mathcal{D}_\kappa {\mathcal{T}^\beta}_{\lambda\beta} - {\mathcal{T}^\gamma}_{\kappa\gamma} {\mathcal{T}^\beta}_{\lambda\beta}) = T_{\mu\nu}.
    \label{A3}
\end{equation}
At this point one ought to recognise that $T_{\mu\nu}$ is the same energy-momentum tensor that describes the electromagnetic field of the Reissner-Nordström spacetime in its GR treatment. This is because $T_{\mu\nu}$ results from the variation of the electromagnetic Lagrangian with respect to the metric $g_{\mu\nu}$, which should for no sensible reason differ once we move to a Finslerian perspective.

The definition of the curvature tensor in a Finsler spacetime guarantees that it reduces to its Riemannian counterpart as long as the metric has no dependence on $y$ and thus all derivatives acting on it reduce to partial derivatives. It therefore satisfies, for the Reissner-Nordström metric, $R = 0$ and consequently $R_{\mu\nu} = T_{\mu\nu}$. We are thus left with
\begin{equation}\label{reducedfe}
    (\delta^{(\lambda}_\nu \delta^{\kappa )}_\mu - g^{\kappa\lambda}g_{\mu\nu})(\mathcal{D}_\kappa {\mathcal{T}^\beta}_{\lambda\beta} - {\mathcal{T}^\gamma}_{\kappa\gamma} {\mathcal{T}^\beta}_{\lambda\beta}) = 0.
\end{equation}
The torsion terms $${\mathcal{T}^\gamma}_{\kappa\gamma} {\mathcal{T}^\beta}_{\lambda\beta}$$ are neglected, as they are of first order on $w_{\alpha\beta}$, as demonstrated if we substitute \eqref{metric d-connection 2} and the non-linear connection  \begin{equation}
		N^\alpha_\mu = \frac{1}{2} y^\gamma g^{\alpha\beta}\partial_\mu g_{\beta\gamma}
	\end{equation}
into relation \eqref{torsion} taking: 
\begin{equation}
    {\mathcal{T}^\alpha}_{\mu\alpha}=-\frac{1}{2}\delta_{\mu}w
    \label{A6}
\end{equation}
with $w=g_{\alpha\beta}w^{\alpha\beta}$. By taking the trace on the remaining terms of \eqref{A3} we obtain the following equation which we will then substitute in the same relation 
\begin{equation}
    g^{\mu\nu} {\mathcal{D}_\mu} {\mathcal{T}^{\alpha}}_{\nu\alpha}= T_{\mu\nu}
\end{equation}
concluding that 
\begin{equation}
\mathcal{D}_{(\mu}{\mathcal{T}^{\alpha}}_{\nu)\alpha}= T_{\mu\nu}-R_{\mu\nu}
\label{A8}
\end{equation}

Proceeding with the application of the non-vanishing components of the linear connection (d-connection) \eqref{D delta nu}, \eqref{D partial b} we receive the definition for partial covariant differentiation as seen below e.g for $X\in TTM$:
\begin{equation}
X^A_{\mid \nu} \equiv D_\nu X^A = \delta_\nu X^A + L^A_{\nu B} X^B
\end{equation}
which is the h-derivative. In view of this, equation \eqref{A8} becomes 
\begin{equation}
\delta_{(\mu} \mathcal{T}^\alpha_{\gamma) \alpha} - L^K_{\mu \nu} \mathcal{T}^\alpha_{K \alpha} = T_{\mu\nu}-R_{\mu\nu} .
\label{A10}
\end{equation}

Returning to equation \eqref{A6}, we proceed by substituting $w$ with its explicit analytical expression,
\begin{equation}
w_{\alpha\beta} = \frac{1}{\sigma} \left( A_{\beta} g_{\gamma\beta} y^{\gamma} + A_{\gamma} g_{\alpha\beta} y^{\gamma} \right) + \frac{1}{{\sigma}^3} A_{\gamma} \delta_{\alpha\epsilon} g_{\beta\eta} y^{\gamma} y^{\epsilon} y^{\eta}
\end{equation}
we obtain ${\mathcal{T}^\alpha}_{\mu\alpha}$ in terms of $A_{\gamma}$:
\begin{equation}
    {\mathcal{T}^\alpha}_{\mu\alpha}=-\frac{5}{2} \delta_{\mu} \left(A_{\gamma}\frac{y^{\gamma}}{\sigma} \right)
    \label{A12}
\end{equation}
Moreover, we can show that \begin{equation}
    \delta_{\nu}\left(\frac{y^{\beta}}{\sigma}\right)= - E^{\beta}_{\gamma\nu}\frac{y^{\gamma}}{\sigma}
    \label{A13}
\end{equation}
with \begin{equation}
    E^{\beta}_{\gamma\nu} (x)\equiv \frac{1}{2}g^{\alpha\beta}\partial_{\nu}g_{\alpha\gamma}
\end{equation}
Using   \eqref{A12} and \eqref{A13} we finally acquire
\begin{equation}
    {\mathcal{T}^\alpha}_{\mu\alpha}=-\frac{5}{2} {\mathcal{K}_{\gamma\nu} \frac{y^{\gamma}}{\sigma}}
    \label{A15}
\end{equation}
where \begin{equation}
    {\mathcal{K}}_{\gamma\nu}(x)\equiv \partial_{\nu}A_{\gamma}- A_{\beta}E^{\beta}_{\gamma\nu}
\end{equation}
From relation \eqref{A15} we can calculate further \eqref{A10} \begin{equation}
    \left[ \frac{1}{2}\partial_{(\mu}\left(\frac{1}{\sigma}\mathcal{K}_{\nu)\gamma}\right)-\frac{1}{2\sigma}\mathcal{L}^{\kappa}_{\mu\nu}\mathcal{K}_{\gamma\nu}  \right]y^{\gamma}=\frac{1}{5}\left(R_{\mu\nu}-T_{\mu\nu}\right)
    \label{A17}
\end{equation}
Relation \eqref{A17} must be satisfied for every choice of $y$. Since the expression enclosed within parentheses does not exhibit any dependence on $y$, we are compelled to conclude that the expression must vanish identically to hold the relation universally.



\begin{thebibliography}{999}

\bibitem{CANTATA:2021asi}
E.~N.~Saridakis \textit{et al.} [CANTATA],
 Modified Gravity and Cosmology. An Update by the CANTATA Network,
Springer, 2021,
[arXiv:2105.12582 [gr-qc]].


\bibitem{DiValentino:2025sru}
Di Valentino, E.; Levi Said, J.; Riess, A.; Pollo, A.; Poulin, V.; Gómez-Valent, A.; Weltman, A.; Palmese, A.; Huang, C.D.; van de Bruck, C.; et al. The CosmoVerse White Paper: Addressing observational tensions in cosmology with systematics and fundamental physics. \textit{arXiv} \textbf{2025}, arXiv.2504.01669. \url{https://doi.org/10.48550/arXiv.2504.01669}.

\bibitem{Shao2024}
Shao, Y.; et al. Semi-Dirac fermions in a topological metal. \textit{Phys. Rev. X} \textbf{2024}, \textit{14}, 041057. \url{https://doi.org/10.1103/PhysRevX.14.041057}.

\bibitem{campanelli2011}
Campanelli, L.; Cea, P.; Fogli, G.; Tedesco, L. Cosmic parallax in ellipsoidal universe. \textit{Modern Physics Letters A} \textbf{2011}, \textit{26}, 1169-1181 . \url{https://doi.org/10.1142/S0217732311035638}.

\bibitem{campanelli2006}
Campanelli, L.; Cea, P.; Tedesco, L. Ellipsoidal universe can solve the cosmic microwave background quadrupole problem. \textit{Phys. Rev. Lett.} \textbf{2006}, \textit{97}, 131302. \url{https://doi.org/10.1103/PhysRevLett.97.131302}.

\bibitem{campanelli2007}
Campanelli, L.; Cea, P.; Tedesco, L. Cosmic microwave background quadrupole and ellipsoidal universe. \textit{Physical Review D} \textbf{2007}, \textit{76}, 063007. \url{https://doi.org/10.1103/PhysRevD.76.063007}.

\bibitem{shamir2025}
Shamir, L. The distribution of galaxy rotation in JWST Advanced Deep Extragalactic Survey. \textit{Mon. Not. R. Astron. Soc.} \textbf{2025}, \textit{538}, 76--91. \url{https://doi.org/10.1093/mnras/staf292}.

\bibitem{MTW}
Misner, C.W.; Thorne, K.S.; Wheeler, J.A. \textit{Gravitation}; Princeton University Press: Princeton, NJ, USA, 2017; ISBN 978-0-691-17779-3.

\bibitem{migkas2021}
Migkas, K.; Pacaud, F.; Schellenberger, G.; Erler, J.; Nguyen-Dang, N.T.; Reiprich, T.H.; Ramos-Ceja, M.E.; Lovisari, L. Cosmological implications of the anisotropy of ten galaxy cluster scaling relations. \textit{Astronomy \& Astrophysics} \textbf{2021}, \textit{649}, A151. \url{https://doi.org/10.1051/0004-6361/202140296}.

\bibitem{pandya2021}
Pandya, A.; Migkas, K.; Reiprich, T.H.; Stanford, A.; Pacaud, F.; Schellenberger, G.; Lovisari, L.; Ramos-Ceja, M.E.; Nguyen-Dang, N.T.; Park, S. Examining the local Universe isotropy with galaxy cluster velocity dispersion scaling relations. \textit{Astronomy \& Astrophysics} \textbf{2021}, \textit{691}, A355. \url{https://doi.org/10.1051/0004-6361/202451755}.

\bibitem{tedesco2024}
Tedesco, L. Ellipsoidal Universe and Cosmic Shear. \textit{Universe} \textbf{2024}, \textit{10}, 363. \url{https://doi.org/10.3390/universe10090363}.

\bibitem{triantafyllopoulos2020}
Triantafyllopoulos, A.; Basilakos, S.; Kapsabelis, E.; Stavrinos, P.C. Schwarzschild-like solutions in Finsler-Randers gravity. \textit{The European Physical Journal C} \textbf{2020}, \textit{80}, 1200. \url{https://doi.org/10.1140/epjc/s10052-020-08772-4}.

\bibitem{kapsabelis2021}
Kapsabelis, E.; Triantafyllopoulos, A.; Basilakos, S.; Stavrinos, P.C. Applications of the Schwarzschild-Finsler-Randers model. \textit{The European Physical Journal C} \textbf{2021}, \textit{81}, 990. \url{https://doi.org/10.1140/epjc/s10052-021-09790-6}.


\bibitem{Stavrinos2004}
Stavrinos, P.; Diakogiannis, F. Finslerian Structure of Anisotropic Gravitational Field. \textit{Gravitation and Cosmology} \textbf{2004}, \textit{10}, 269--278.

\bibitem{Vacaru2005}
Vacaru, S.; Stavrinos, P.; Gabourov, E.; Gontsa, D. Clifford and Riemann-Finsler Structures in Geometric Mechanics and Gravity. Eds. \textit{Geometry Balkan Press} \textbf{2005}, pp. 643. arXiv:gr-qc/0508023. \url{https://arxiv.org/abs/gr-qc/0508023}.

\bibitem{Kouretsis2009}
Kouretsis, A.P.; Stathakopoulos, M.; Stavrinos, P.C. The General Very Special Relativity in Finsler Cosmology. \textit{Phys. Rev. D} \textbf{2009}, \textit{79}, 104011. \url{https://doi.org/10.1103/PhysRevD.79.104011}.

\bibitem{Skakala2011}
Skakala, J.; Visser, M. Bi-metric pseudo-Finslerian spacetimes. \textit{J. Geom. Phys.} \textbf{2011}, \textit{61}, 1396--1400. \url{https://doi.org/10.1016/j.geomphys.2011.03.003}.

\bibitem{Kostelecky2011}
Kostelecky, A. Riemann-Finsler geometry and Lorentz-violating kinematics. \textit{Phys. Lett. B} \textbf{2011}, \textit{701}, 137--143. \url{https://doi.org/10.1016/j.physletb.2011.05.041}.

\bibitem{Vacaru2012}
Vacaru, S.I. Principles of Einstein-Finsler Gravity and Perspectives in Modern Cosmology. \textit{Int. J. Mod. Phys. D} \textbf{2012}, \textit{21}, 1250072. \url{https://doi.org/10.1142/S0218271812500721}.

\bibitem{Pfeifer2012}
Pfeifer, C.; Wohlfarth, M.N.R. Finsler geometric extension of Einstein gravity. \textit{Phys. Rev. D} \textbf{2012}, \textit{85}, 064009. \url{https://doi.org/10.1103/PhysRevD.85.064009}.


 
\bibitem{Basilakos:2013hua}
S.~Basilakos, A.~P.~Kouretsis, E.~N.~Saridakis and P.~Stavrinos,
 Resembling dark energy and modified gravity with Finsler-Randers cosmology, 
Phys. Rev. D \textbf{88}, 123510 (2013)
 \url{https://doi.org/10.1103/PhysRevD.88.123510}.
 


\bibitem{Pfeifer2011}
Pfeifer, C.; Wohlfarth, M.N.R. Causal structure and electrodynamics on Finsler spacetimes. \textit{Phys. Rev. D} \textbf{2011}, \textit{84}, 044039. \url{https://doi.org/10.1103/PhysRevD.84.044039}.

\bibitem{Elizalde2015}
Elizalde, E.; Vacaru, S.I. Einstein spaces modeling nonminimal modified gravity theories. \textit{Eur. Phys. J. Plus} \textbf{2015}, \textit{130}, 119. \url{https://doi.org/10.1140/epjp/i2015-15119-0}.

\bibitem{Caponio2016}
Caponio, E.; Stancarone, G. Standard static Finsler spacetimes. \textit{Int. J. Geom. Methods Mod. Phys.} \textbf{2016}, \textit{30}, 1650040. \url{https://doi.org/10.1142/S0219887816500407}.

\bibitem{Brody2016}
Brody, D.C.; Gibbons, G.W.; Meier, D.M.A Riemannian approach to Randers geodesics. \textit{J. Geom. Phys.} \textbf{2016}, \textit{106}, 98--101. \url{https://doi.org/10.1016/j.geomphys.2016.03.019}.

\bibitem{Hohmann2017}
Hohmann, M.; Pfeifer, C. Geodesics and the magnitude-redshift relation in general Finsler spacetimes. \textit{Phys. Rev. D} \textbf{2017}, \textit{95}, 104021. \url{https://doi.org/10.1103/PhysRevD.95.104021}.

\bibitem{WangEPJC2017}
Wang, D.; Meng, X.-H. Seeking sterile neutrinos in Finslerian cosmology. \textit{Eur. Phys. J. C} \textbf{2017}, \textit{77}, 725. \url{https://doi.org/10.1140/epjc/s10052-017-5284-9}.

\bibitem{Ellis2017}
Ellis, J.; Mavromatos, N.E.; Nanopoulos, D.V. Novel foamy origin for singlet fermion masses. \textit{Phys. Rev. D} \textbf{2017}, \textit{96}, 086012. \url{https://doi.org/10.1103/PhysRevD.96.086012}.

\bibitem{Silva2016}
Silva, J.E.G. Symmetries and fields in Randers-Finsler spacetime. \textit{arXiv} \textbf{2016}, arXiv:1602.07345. \url{https://arxiv.org/abs/1602.07345}.

\bibitem{Wang2017}
Wang, D.; Meng, X.-H. Finslerian universe may reconcile tensions between high and low redshift probes. \textit{arXiv} \textbf{2017}, arXiv:1709.04141. \url{https://arxiv.org/abs/1709.04141}.

\bibitem{Shah2018}
Shah, M.B.; Ganai, P.A. Quantum gauge freedom in Lorentz violating background: A Finsler geometry approach. \textit{Int. J. Geom. Methods Mod. Phys.} \textbf{2018}, \textit{15}, 1850009. \url{https://doi.org/10.1142/S0219887818500093}.

\bibitem{Fuster2018}
Fuster, A.; Pabst, C.; Pfeifer, C. Berwald spacetimes and very special relativity. \textit{Phys. Rev. D} \textbf{2018}, \textit{98}, 084062. \url{https://doi.org/10.1103/PhysRevD.98.084062}.

\bibitem{Triantafyllopoulos2018}
Triantafyllopoulos, A.; Stavrinos, P.C. Weak field equations and generalized FRW cosmology on the tangent Lorentz bundle. \textit{Class. Quantum Grav.} \textbf{2018}, \textit{35}, 085011. \url{https://doi.org/10.1088/1361-6382/aab27f}.

\bibitem{Chaubey2019}
Chaubey, R.; Pradhan, A.; Mishra, B.; Rani, S. Finsler-Randers Cosmological Models in the Presence of Massive String Cloud with Magnetic Field. \textit{Proc. Natl. Interdisc. Symp. Innov.} \textbf{2019}.
\bibitem{Minas:2019urp}
G.~Minas, E.~N.~Saridakis, P.~C.~Stavrinos and A.~Triantafyllopoulos,
 Bounce cosmology in generalized modified gravities, 
Universe \textbf{5}, 74 (2019)
\url{https://doi.org/10.3390/universe5030074}.

 
\bibitem{Bubuianu2019}
Bubuianu, L.; Vacaru, S.I. Black holes with modified dispersion relations and solitonic configurations in Einstein-Finsler gravity. \textit{Ann. Phys.} \textbf{2019}, \textit{404}, 10--38. \url{https://doi.org/10.1016/j.aop.2019.02.013}.

 
\bibitem{Ikeda:2019ckp}
S.~Ikeda, E.~N.~Saridakis, P.~C.~Stavrinos and A.~Triantafyllopoulos,
 Cosmology of Lorentz fiber-bundle induced scalar-tensor theories, 
Phys. Rev. D \textbf{100}, no.12, 124035 (2019)
\url{https://doi.org/10.1103/PhysRevD.100.124035}.
 

\bibitem{HohmannU2020}
Hohmann, M.; Pfeifer, C.; Voicu, N. Cosmological Finsler Spacetimes. \textit{Universe} \textbf{2020}, \textit{6}, 65. \url{https://doi.org/10.3390/universe6050065}.
 
\bibitem{Caponio2020}
Caponio, E.; Masiello, A. On the Analyticity of Static Solutions of a Field Equation in Finsler Gravity. \textit{Universe} \textbf{2020}, \textit{6}, 59. \url{https://doi.org/10.3390/universe6040059}.

\bibitem{Relancio2020}
Relancio, J.J.; Liberati, S. Phenomenological consequences of a geometry in the cotangent bundle. \textit{Phys. Rev. D} \textbf{2020}, \textit{101}, 064062. \url{https://doi.org/10.1103/PhysRevD.101.064062}.

\bibitem{HohmannEPJC2020}
Hohmann, M.; Pfeifer, C.; Voicu, N. The kinetic gas universe. \textit{Eur. Phys. J. C} \textbf{2020}, \textit{80}, 809. \url{https://doi.org/10.1140/epjc/s10052-020-8324-z}.

\bibitem{HohmannPRD2020}
Hohmann, M.; Pfeifer, C.; Voicu, N. Relativistic kinetic gases as direct sources of gravity. \textit{Phys. Rev. D} \textbf{2020}, \textit{101}, 024062. \url{https://doi.org/10.1103/PhysRevD.101.024062}.

\bibitem{Javaloyes2021}
Javaloyes, M. Á.; Sánchez, M.; Villaseñor, F.F. The Einstein-Hilbert-Palatini formalism in pseudo-Finsler geometry. \textit{Adv. Theor. Math. Phys.} \textbf{2022}, \textit{26}, 10. \url{https://dx.doi.org/10.4310/ATMP.2022.v26.n10.a5}.

 
\bibitem{Konitopoulos:2021eav}
S.~Konitopoulos, E.~N.~Saridakis, P.~C.~Stavrinos and A.~Triantafyllopoulos,
 Dark gravitational sectors on a generalized scalar-tensor vector bundle model and cosmological applications,''\\
Phys. Rev. D \textbf{104}, no.6, 064018 (2021)
 \url{https://dx.doi.org/10.1103/PhysRevD.104.064018}.
 


 
\bibitem{vanVoorthuizen2021}
van Voorthuizen, J. Kasner metric in Finsler gravity. \textit{TUE Research portal} \textbf{2021}, Bachelor thesis.

\bibitem{Hama2021}
Hama, R.; Harko, T.; Sabau, S.V.; Shahidi, S. Cosmological evolution and dark energy in osculating Barthel-Randers geometry. \textit{Eur. Phys. J. C} \textbf{2021}, \textit{81}, 742. \url{https://doi.org/10.1140/epjc/s10052-021-09517-7}.

\bibitem{Stavrinos2021}
Stavrinos, P.; Vacaru, S.I. Broken Scale Invariance, Gravity Mass, and Dark Energy in Modified Einstein Gravity with Two Measure Finsler Like Variables. \textit{Universe} \textbf{2021}, \textit{7}, 89. \url{https://doi.org/10.3390/universe7040089}.

\bibitem{Hama2022}
Hama, R.; Harko, T.; Sabau, S.V. Dark energy and accelerating cosmological evolution from osculating Barthel-Kropina geometry. \textit{Eur. Phys. J. C} \textbf{2022}, \textit{82}, 385. \url{https://doi.org/10.1140/epjc/s10052-022-10318-9}.

\bibitem{Hama2023}
Hama, R.; Harko, T.; Sabau, S.V. Conformal gravitational theories in Barthel-Kropina-type Finslerian geometry, and their cosmological implications. \textit{Eur. Phys. J. C} \textbf{2023}, \textit{83}, 1030. \url{https://doi.org/10.1140/epjc/s10052-023-12146-x}.

\bibitem{Heefer2023}
Heefer, S.; Pfeifer, C.; Reggio, A.; Fuster, A. A Cosmological unicorn solution to Finsler gravity. \textit{Phys. Rev. D} \textbf{2023}, \textit{108}, 064051. \url{https://doi.org/10.1103/PhysRevD.108.064051}.

\bibitem{Narasimhamurthy2024}
Narasimhamurthy, S.K.; Praveen, J.
Cosmological constant roll of inflation within Finsler-barthel-Kropina geometry: A geometric approach to early universe dynamics.
\textit{New Astronomy} \textbf{2024}, \textit{108}, 102187.
\url{https://doi.org/10.1016/j.newast.2024.102187}.

\bibitem{AnnamáriaFriedl-Szász2025}
Friedl-Szász, A.; Popovici-Popescu, E.; Voicu, N.; Pfeifer, C.; Heefer, S. Cosmological Landsberg-Finsler spacetimes. \textit{Phys. Rev. D} \textbf{2025}, \textit{111}, 044058. \url{https://doi.org/10.1103/PhysRevD.111.044058}.

\bibitem{Zhu2023}
Zhu, J.; Ma, B.-Q. Lorentz violation in Finsler geometry. \textit{Symmetry} \textbf{2023}, \textit{15}, 978. \url{https://doi.org/10.3390/sym15050978}.

\bibitem{Savvopoulos2023}
Savvopoulos, C.; Stavrinos, P.C. Anisotropic conformal dark gravity on the Lorentz tangent bundle spacetime. \textit{Phys. Rev. D} \textbf{2023}, \textit{108}, 044048. \url{https://doi.org/10.1103/PhysRevD.108.044048}.

\bibitem{Villasenor2024}
Fernández Villaseñor, F. Lorentz-Finsler geometry and Einstein equations. Ph.D. Dissertation. \textit{University of Granada (joint with U.’s of
Almería, Cádiz, Jaén and Málaga)} \textbf{2024}.

\bibitem{Randers1941}
Randers, G. On an Asymmetrical Metric in the Four-Space of General Relativity. \textit{Phys. Rev.} \textbf{1941}, \textit{59}, 195. \url{https://doi.org/10.1103/PhysRev.59.195}.

\bibitem{Gibbons2007}
Gibbons, G.W.; Gomis, J.; Pope, C.N. General very special relativity is Finsler geometry. \textit{Phys. Rev. D} \textbf{2007}, \textit{76}, 081701. \url{https://doi.org/10.1103/PhysRevD.76.081701}.

\bibitem{Chaubey2018}
Chaubey, R.; Tiwari, B.; Shukla, A.; Kumar, M.
Finsler–Randers Cosmological Models in Modified Gravity Theories.
\textit{Proc. Nat. Inst. Sci. India (Pt. A Phys. Sci.)} \textbf{2019}, \textit{89}, 757--768.
\url{https://doi.org/10.1007/s40010-018-0534-2}.

 



\bibitem{Raushan2020}
Raushan, R.; Chaubey, R. Finsler-Randers cosmology in the framework of a particle creation mechanism: a dynamical systems perspective. \textit{Eur. Phys. J. Plus} \textbf{2020}, \textit{135}, 228. \url{https://doi.org/10.1140/epjp/s13360-020-00221-1}.

\bibitem{Mavromatos2013}
Mavromatos, N.E. Violation of CPT invariance in the Early Universe: Strings in a Robertson-Walker background and particle-antiparticle asymmetries. \textit{J. Phys. Conf. Ser.} \textbf{2013}, \textit{447}, 012016. \url{https://doi.org/10.1088/1742-6596/447/1/012016}.

\bibitem{Basilakos2013}
Basilakos, S.; Stavrinos, P. Cosmological equivalence between the Finsler-Randers space-time and the DGP gravity model. \textit{Phys. Rev. D} \textbf{2013}, \textit{87}, 043506. \url{https://doi.org/10.1103/PhysRevD.87.043506}.

\bibitem{Vacaru2014}
Vacaru, S.I. Exact solutions in modified massive gravity and off-diagonal wormhole deformations. \textit{Eur. Phys. J. C} \textbf{2014}, \textit{74}, 2781. \url{https://doi.org/10.1140/epjc/s10052-014-3132-3}.

\bibitem{Vacaru2017}
Vacaru, S.I.; Irwin, K. Off-diagonal deformations of Kerr metrics and black ellipsoids in heterotic supergravity. \textit{Eur. Phys. J. C} \textbf{2017}, \textit{77}, 17. \url{https://doi.org/10.1140/epjc/s10052-014-2781-y}.

\bibitem{Caponio2018}
Caponio, E.; Stancarone, G. On Finsler spacetimes with a timelike Killing vector field. \textit{Class. Quantum Grav.} \textbf{2018}, \textit{35}, 085007. \url{https://doi.org/10.1088/1361-6382/aab0d9}.

\bibitem{Stavrinos2018}
Stavrinos, P.C.; Alexiou, M. Raychaudhuri equation in the Finsler-Randers space-time and generalized scalar-tensor theories. \textit{Int. J. Geom. Methods Mod. Phys.} \textbf{2018}, \textit{15}, 1850039. \url{https://doi.org/10.1142/S0219887818500391}.

\bibitem{Heefer2020}
Heefer, S.; Pfeifer, C.; Fuster, A. Randers pp-waves. \textit{Phys. Rev. D} \textbf{2021}, \textit{104}, 024007. \url{https://doi.org/10.1103/PhysRevD.104.024007}.

\bibitem{TriantafyllopoulosEPJC2020}
Triantafyllopoulos, A.; Basilakos, S.; Kapsabelis, E.; Stavrinos, P.C. Schwarzschild-like solutions in Finsler-Randers gravity. \textit{Eur. Phys. J. C} \textbf{2020}, \textit{80}, 1200. \url{https://doi.org/10.1140/epjc/s10052-020-08772-4}.

\bibitem{Elbistan2020}
Elbistan, M.; Zhang, P.M.; Dimakis, N.; Gibbons, G.W.; Horvathy, P.A. Geodesic motion in Bogoslovsky-Finsler spacetimes. \textit{Phys. Rev. D} \textbf{2020}, \textit{102}, 024014. \url{https://doi.org/10.1103/PhysRevD.102.024014}.

\bibitem{Silva2021}
Silva, J.E.G. A field theory in Randers-Finsler spacetime. \textit{EPL} \textbf{2021}, \textit{133}, 21002. \url{https://doi.org/10.1209/0295-5075/133/21002}.

\bibitem{Angit2022}
Angit, S.; Raushan, R.; Chaubey, R. Stability and bifurcation analysis of Finsler-Randers cosmological model. \textit{Pramana} \textbf{2022} \textit{96}, 123. \url{https://doi.org/10.1007/s12043-022-02363-6}.

\bibitem{Lou2022}
Lou, H.; Li, J.; Yang, W.; Feng, W.; Liu, W.; Zhang, Q.; Zhang, N.; Qi, Y.; Wu, Y. Theoretical analysis on the Rényi holographic dark energy in the Finsler-Randers cosmology. \textit{Int. J. Mod. Phys. D} \textbf{2022}, \textit{31}, 2250002. \url{https://doi.org/10.1142/S021827182250002X}.

\bibitem{Kapsabelis2022}
Kapsabelis, E.; Kevrekidis, P.G.; Stavrinos, P.C.; Triantafyllopoulos, A. Schwarzschild-Finsler-Randers spacetime: geodesics, dynamical analysis and deflection angle. \textit{Eur. Phys. J. C} \textbf{2022}, \textit{82}, 1908. \url{https://doi.org/10.1140/epjc/s10052-022-11081-7}.

\bibitem{El-Nabulsi2024}
El-Nabulsi, A.R.; Golmankhaneh, A.K. Nonstandard and fractal electrodynamics in Finsler-Randers space. \textit{Int. J. Geom. Methods Mod. Phys.} \textbf{2022}, \textit{19}, 2250080. \url{https://doi.org/10.1142/S0219887822500803}.

\bibitem{Voicu2023}
Voicu, N.; Friedl-Szász, A.; Popovici-Popescu, E.; Pfeifer, C. The Finsler spacetime condition for $(\alpha, \beta)$-metrics and their isometries. \textit{Universe} \textbf{2023}, \textit{9}, 198. \url{https://doi.org/10.3390/universe9040198}.

\bibitem{Feng2023}
Feng, W.; Lou, H.; Li, X. Theoretical analysis on the Barrow holographic dark energy in the Finsler-Randers cosmology. \textit{Int. J. Mod. Phys. D} \textbf{2023}, \textit{32}, 2350029. \url{https://doi.org/10.1142/S0218271823500293}.

\bibitem{Das2023}
Das, K.P.; Debnath, U. Possible existence of traversable wormhole in Finsler-Randers geometry. \textit{Eur. Phys. J. C} \textbf{2023}, \textit{83}, 821. \url{https://doi.org/10.1140/epjc/s10052-023-11910-3}.

\bibitem{Kapsabelis2024}
Kapsabelis, E.; Saridakis, E.N.; Stavrinos, P.C. Finsler-Randers-Sasaki gravity and cosmology. \textit{Eur. Phys. J. C} \textbf{2024}, \textit{84}, 538. \url{https://doi.org/10.1140/epjc/s10052-024-12924-1}.

\bibitem{Chanda2024}
Chanda, S. More on Jacobi metric: Randers-Finsler metrics, frame dragging and geometrisation techniques. \textit{Eur. Phys. J. Plus} \textbf{2024}, \textit{139}, 983. \url{https://doi.org/10.1140/epjp/s13360-024-05775-y}.

\bibitem{Silagadze2010}
Silagadze, Z.K. On the Finslerian extension of the Schwarzschild metric. \textit{Acta Phys. Polon. B} \textbf{2010}, \textit{41}, 2171--2176. \url{https://doi.org/10.5506/APhysPolB.42.1199}.

\bibitem{Li2020}
Li, X.; Zhao, S.-P. Quasinormal modes in Finslerian-Schwarzschild spacetime. \textit{Phys. Rev. D} \textbf{2020}, \textit{101}, 124012. \url{https://doi.org/10.1103/PhysRevD.101.124012}.

\bibitem{Kapsabelis2021}
Kapsabelis, E.; Basilakos, S.; Triantafyllopoulos, A.; Stavrinos, P. Applications of Schwarzschild-Finsler-Randers model. \textit{Eur. Phys. J. C} \textbf{2021}, \textit{81}, 931. \url{https://doi.org/10.1140/epjc/s10052-021-09790-6}.

\bibitem{Ke-JianHe2024}
He, K.-J.; Yao, J.-T.; Zhang, X.; Li, X. Shadows and photon motions in axially symmetric Finslerian Schwarzschild black holes. \textit{Phys. Rev. D} \textbf{2024}, \textit{109}, 064049. \url{https://doi.org/10.1103/PhysRevD.109.064049}.

\bibitem{Dehkordi2025}
Dehkordi, H.R.; Richartz, M.; Saa, A. Finslerian structure of black holes. \textit{arXiv} \textbf{2025}, \url{https://doi.org/10.48550/arXiv.2501.09536}.

\bibitem{Yao2025}
Yao, J.-T.; Zhu, Q.-H.; Li, X.
Identifying axially symmetric Finslerian extensions of Schwarzschild black hole via the S2 star orbiting Sagittarius $A^\star$.
\textit{Phys. Rev. D} \textbf{2025}, \textit{111}, 084083.
\url{https://doi.org/10.1103/PhysRevD.111.084083}.

\bibitem{Konoplya:2006br}
Konoplya, R.A.; Zhidenko, A. Stability and quasinormal modes of the massive scalar field around Kerr black holes. \textit{Phys. Rev. D} \textbf{2006}, \textit{73}, 124040. \url{https://doi.org/10.1103/PhysRevD.73.124040}.

\bibitem{Rajpoot2015}
Rajpoot, S.; Vacaru, S.I. Black Ring and Kerr Ellipsoid—Solitonic Configurations in Modified Finsler Gravity. \textit{Int. J. Geom. Methods Mod. Phys.} \textbf{2015}, \textit{12}, 1550102. \url{https://doi.org/10.1142/S0219887815501029}.

\bibitem{Li2018}
Li, X. Special Finslerian generalization of the Reissner-Nordström spacetime. \textit{Phys. Rev. D} \textbf{2018}, \textit{98}, 084030. \url{https://doi.org/10.1103/PhysRevD.98.084030}.

\bibitem{Rarras2024}
Rarras, D.; Kosmas, T.; Papavasileiou, T.; Kosmas, O. Galactic Stellar Black Hole Binaries: Spin Effects on Jet Emissions of High-Energy Gamma-Rays. \textit{Particles} \textbf{2024}, \textit{7}, 792--804. \url{https://doi.org/10.3390/particles7030046}.

\bibitem{Kerr1963}
Kerr, R.P. Gravitational field of a spinning mass as an example of algebraically special metrics. \textit{Phys. Rev. Lett.} \textbf{1963}, \textit{11}, 237--238. \url{https://doi.org/10.1103/PhysRevLett.11.237}.

\bibitem{Kerr1970}
Kerr, R.P.; Debney, G.C. Einstein spaces with symmetry groups. \textit{J. Math. Phys.} \textbf{1970}, \textit{11}, 2807--2812. \url{https://doi.org/10.1063/1.1665451}.

\bibitem{Visser2007}
Visser, M. The Kerr spacetime: A brief introduction. \textit{arXiv} \textbf{2007}, arXiv:0706.0622. \url{https://doi.org/10.48550/arXiv.0706.0622}.

\bibitem{Stavrinos2008}
Stavrinos, P.C.; Kouretsis, A.P.; Stathakopoulos, M. Friedman-like Robertson-Walker model in generalized metric space-time with weak anisotropy. \textit{Gen. Relativ. Gravit.} \textbf{2008}, \textit{40}, 1403--1425. \url{https://doi.org/10.1007/s10714-007-0540-1}.

\bibitem{Cvetic2014}
Cvetič, M.; Gibbons, G.W. Exact quasinormal modes for the near horizon Kerr metric. \textit{Phys. Rev. D} \textbf{2014}, \textit{89}, 064057. \url{https://doi.org/10.1103/PhysRevD.89.064057}.


\bibitem{Kerr2023}
Kerr, R.P. Do Black Holes have Singularities? \textit{arXiv} \textbf{2023}, arXiv.2312.00841. \url{https://doi.org/10.48550/arXiv.2312.00841}.

\bibitem{Miron1994}
Miron, R.; Anastasiei, M. The Geometry of Lagrange Spaces: Theory and Applications. In \textit{Fundamental Theories of Physics}; Springer:
Heidelberg, The Netherlands, 1994. \url{https://dx.doi.org/10.1007/978-94-011-0788-4}.

\bibitem{Ahluwalia2010}
Ahluwalia, D.V.; Horvath, S.P. Very special relativity as relativity of dark matter: The Elko connection. \textit{J. High Energy Phys.} \textbf{2010}, \textit{2010}, 78. \url{https://doi.org/10.1007/JHEP11(2010)078}.

\bibitem{Capozziello2011}
Capozziello, S.; De Laurentis, M. Extended Theories of Gravity. \textit{Phys. Rep.} \textbf{2011}, \textit{509}, 167--321. \url{https://doi.org/10.1016/j.physrep.2011.09.003}.

\bibitem{Silva2017}
Silva, J.E.G.; Maluf, R.V.; Almeida, C.A.S. A nonlinear dynamics for the scalar field in Randers spacetime. \textit{Phys. Lett. B} \textbf{2017}, \textit{766}, 263-267. \url{https://doi.org/10.1016/j.physletb.2017.01.025}.

\bibitem{Chirco2022}
Chirco, G.; Liberati, S.; Relancio, J.J. Spacetime thermodynamics in momentum-dependent geometries. \textit{Phys. Rev. D} \textbf{2022}, \textit{106}, 064048. \url{https://doi.org/10.1103/PhysRevD.106.064048}.

\bibitem{Szigeti2025}
Szigeti, B.E.; Szapudi, I.; Barna, I.F.; Barnaföldi, G.G. Can rotation solve the Hubble Puzzle? \textit{Mon. Not. Roy. Astron. Soc.} \textbf{2025}, \textit{538}, 3038--3041. \url{https://doi.org/10.1093/mnras/staf446}.

\bibitem{Koivisto2008}
Koivisto, T.; Mota, D.F. Accelerating cosmologies with an anisotropic equation of state. \textit{Astrophys. J.} \textbf{2008}, \textit{679}, 1. \url{https://doi.org/10.1086/587537}.

\bibitem{Vacaru2001}
Vacaru, S.; Stavrinos, P. Spinors and Space-Time Anisotropy. \textit{arXiv} \textbf{2001}, arXiv:gr-qc/0112028. \url{https://doi.org/10.48550/arXiv.gr-qc/0112028}.

\bibitem{rarras2024}
Rarras, D.; Kosmas, O.; Papavasileiou, T.; Kosmas, T. Black Hole's Spin-Dependence of $\gamma$-Ray and Neutrino Emissions from MAXI J1820+070, XTE J1550-564, and XTE J1859+226. \textit{Particles} \textbf{2024}, \textit{7}, 818--833. \url{https://doi.org/10.3390/particles7030049}.


\end{thebibliography}
\end{document}